\documentclass[aps,showkeys]{revtex4}
\usepackage{amsfonts}
\usepackage{amsmath}
\usepackage{amssymb}
\usepackage{graphicx}

\begin{document}
\title[Brush Effects on DNA Chips]{Brush Effects on 
DNA Chips: Thermodynamics, Kinetics and Design 
Guidlines}
\author{A. Halperin and A. Buhot}
\affiliation{UMR 5819 SPrAM (UJF, CNRS, CEA), DRFMC, 
CEA Grenoble, 17 rue des Martyrs, 38054 Grenoble 
cedex 9, France}
\author{E. B. Zhulina}
\affiliation{Institute of Macromolecular Compounds 
of the Russian Academy of Sciences, 199004 St 
Petersburg, Russia}
\keywords{DNA microarrays, Hybridization isotherm, 
Hybridization rate, Polyelectrolyte brush, Gene 
expression, Mutation detection}

\begin{abstract}
In biology experiments, oligonucleotide microarrays 
are contacted with a solution of long nucleic acid 
(NA) targets. The hybridized probes thus carry long 
tails. When the surface density of the oligonucleotide
probes is high enough, the progress of hybridization 
leads to the formation of a polyelectrolyte brush due 
to mutual crowding of the NA tails. The free energy 
penalty associated with the brush modifies both the 
hybridization isotherms and the rate equations: the 
attainable hybridization is lowered significantly as 
is the hybridization rate. While the equilibrium 
hybridization fraction, $x_{eq}$, is low, the 
hybridization follows a Langmuir type isotherm, 
$x_{eq}/(1-x_{eq}) = c_t K$ where $c_t$ is the
target concentration and $K$ is the equilibrium 
constant smaller than its bulk value by a factor
$(n/N)^{2/5}$ due to wall effects where $n$ and 
$N$ denote the number of bases in the probe and the 
target. At higher $x_{eq}$, when the brush is 
formed, the leading correction is $x_{eq}/(1-x_{eq}) 
= c_t K \exp [ - const' (x_{eq}^{2/3} - x_B^{2/3})]$ 
where $x_B$ corresponds to the onset of the brush 
regime. The denaturation rate constant in the two 
regimes are identical. However, the hybridization 
rate constant in the brush regime is lower, the 
leading correction being $\exp [- const' (x^{2/3} 
- x_B^{2/3})]$. 
\end{abstract}

\maketitle

\section*{INTRODUCTION}

The growing availability of genomic DNA sequences 
enables research on profiles of gene expression, 
single nucleotide polymorphism (SNP) and their role, 
molecular diagnostics for cancer etc. In turn, these 
activities require simultaneous interrogation of a 
given sample for the presence of numerous different 
nucleic acid sequences. DNA microarrays, ``DNA chips'', 
emerged as an important method for such parallel 
analysis (Graves, 1999; Wang, 2000; Lockhart and 
Winzeler, 2000; Heller, 2002). DNA chips function 
by parallel hybridization of labelled nucleic acid 
sequences in the solution, known as targets, to an 
array of nucleic acid probes bound to a surface. 
Numerous identical probes are localized at a small 
area known as ``spot'' or ``probe cell''. The 
composition of the sample is deduced from the label 
intensities of the different spots after the 
hybridization. DNA chips are produced in one of 
two main formats. In cDNA microarrays, long cDNA 
targets are physisorbed onto the substrate while 
in oligonucleotide chips short oligonucleotides are 
chemically bound to the surface via their terminal 
groups. Our theoretical considerations address the 
hybridization behavior, kinetics and thermodynamics, 
of oligonucleotide microarrays when the targets are 
much longer than the probes as is typically the case 
in biology experiments (See for examples: Guo et al., 
1994; Prix et al., 2002). In particular, we analyze 
the consequences of the interactions between the 
long hybridized targets at the surface (Fig.1).

A growing theory effort aims to clarify the underlying 
physics of DNA chips with view of assisting in their 
design and in the analysis of the results. The Langmuir 
isotherm and the corresponding kinetic scheme provide 
a natural starting point for the modeling (Chan et al., 
1995; Livshits and Mirzabekov, 1996; Vainrub and Pettitt, 
2002; Bhanot et al., 2003; Held et al., 2003; Zhang et 
al., 2003; Halperin et al., 2004a, 2004b) as well as the 
analysis of the experimental results (Forman et al., 1998; 
Okahata et al., 1998; Steel et al., 1998; Georgiadis et 
al., 2000; Nelson et al., 2001; Dai et al., 2002; Kepler 
et al., 2002; Peterson et al., 2002; Hekstra et al., 2003). 
Within this model, the probes, irrespective of their 
hybridization state, do not interact. This assumption 
is justified when the probe density in the spots is 
sufficiently low. At higher probe densities interactions 
are no longer negligible and the Langmuir model requires 
modifications. As we shall discuss, the necessary 
modifications depend crucially on the length of the 
targets as characterized by $N$, the number of bases 
or monomers. Importantly, in biology experiments the 
targets are usually significantly longer than the probes. 
As a result, each hybridized probe binds a long segment 
of single stranded nucleic acid formed by the unhybridized 
part of the target (Fig. 1). This leads to two effects. 
First, when the tails do not overlap the hybridization at 
an impenetrable surface incurs an entropic penalty. This 
reduces the equilibrium constant of hybridization with 
repect to its bulk value. Second, it is necessary to allow 
for the crowding of these unhybridized ``tails'' as the 
fraction of hybridized probes grows. This crowding gives 
rise to a polymer brush, a phenomenon that was extensively 
studied in polymer physics (Milner, 1991; Halperin et al., 
1992; R\"uhe et al., 2004). The theory of polyelectrolyte 
brushes (Pincus, 1991; Borisov et al., 1991; R\"uhe et al., 
2004), as modified to allow for target-probe interactions 
and wall effects, enables us to analyze the effects of the 
crowding on the thermodynamics and kinetics of hybridization 
on DNA chips. In particular, we obtain the hybridization 
isotherm and the rate equation in the brush regime when 
the unhybridized tails overlap. As we shall see, the free 
energy penalty associated with the brush gives rise to 
distinctive modification of the Langmuir isotherm and 
kinetics. Importantly, the brush penalty reflects both 
the electrostatic interactions within the probe layer 
and the entropic price due to the extension of the crowded 
chains. It results in slower hybridization and lower 
attainable hybridization.

Our analysis focuses on oligonucleotide microarrays 
hybridizing with long targets of single stranded 
(ss) DNA. For simplicity we limit the discussion to 
the experimentally attainable case of monodispersed 
targets and probes, a passivated surface that eliminates 
physical adsorption of DNA and probes anchored to the 
surface via short spacer chains. The qualitative
features of our results apply however to a wider range 
of systems. Two hybridization regimes appear, depending 
on the equilibrium hybridization fraction, $x_{eq}$, as 
set by the bulk concentration of the target, $c_{t}$. 
For low $x_{eq}$, the hybridization isotherm is of the 
Langmuir form, $x_{eq}/(1-x_{eq}) = c_{t} K$ where $K$ 
is the equilibrium constant of the hybridization reaction 
at the surface. For probes comprising $n \ll N$ bases $K$
at an impenetrable surface is reduced by a factor of 
$(n/N)^{2/5}$ with respect to the equilibrium constant 
of the free chains in solution. At higher $x_{eq}$, 
obtained at higher $c_{t}$, the effective equilibrium 
constant is modified because of the brush penalty. 
The leading correction to the hybridization isotherm 
is $x_{eq}/(1-x_{eq}) =c_{t} K \exp [ -const' 
(x_{eq}^{2/3}-x_{B}^{2/3})]$ where $x_{B}$ corresponds 
to the onset of brush formation. The formation of the 
brush does not affect the denaturation rate constant 
of the hybridized probe. However, it does lower the 
hybridization rate constant by a factor of $\exp[-const'
(x^{2/3}-x_{B}^{2/3})]$ where $x$ is the instantaneous
hybridization fraction. The proportionality constant 
scales with $N/\Sigma_{0}^{2/3}$ where $\Sigma_{0}$ 
is the area per probe.

To our knowledge, there has been no direct experimental 
study of the effects of brush formation on the hybridization 
isotherms and the hybridization rates. Yet, experimental 
evidence of brush effects has been reported. Guo et al. 
(1994) observed that the maximum attainable hybridization 
fraction is reached at higher $\Sigma_{0}$ when $N$ 
increases. Su et al. (2002) reported slower hybridization 
as $N$ increases at fixed $\Sigma_{0}$. A similar effect
was reported for RNA targets by Dai et al. (2002). 
Further support for the existence of the brush effect 
is lent by the wide spread use of sample fragmentation 
to achieve a lower average $N$ (See for example: Rosenow 
et al., 2001; Affymetrix, 2004).

The practical implications of our analysis concern three 
issues: the design of DNA chips, the sample preparation 
and the analysis of the data. The design of DNA chips 
currently reflects the view that an increase in the 
oligonucleotide density in a spot should increase the 
signal intensity and therefore the sensitivity (Pirrung, 
2002). Certain limitations of this strategy, due to the
increase of the DNA diameter upon hybridization and 
the resulting steric hindrance, has been long recognized 
(Southern et al., 1999). In marked distinction, our 
analysis highlights limitations due to the crowding of 
the long non-hybridized tails of the targets. Thus, 
in choosing $\Sigma_{0}$ it is useful to bear in mind 
the anticipated $N$ of the sample and its effect on 
the attainable hybridization. When $\Sigma_{0}$ is 
fixed, our analysis provides guidelines for the sample 
preparation. In particular, the choice of $N$ as 
determined by the PCR primers or the fragmentation 
procedure. Concerning data analysis, our discussion 
identifies possible sources of error when comparing 
spot intensities of samples with different $N$. These 
may occur because both the onset of saturation and 
the hybridization rate vary with $N$. In quantitative 
terms, our analysis yields two guidelines: Concerning 
equilibrium hybridization, it leads to a simple 
relationship between the area per probe, $\Sigma_{0}$, 
the number of bases in the probe, $n$, the number of 
bases in the target, $N$, and the attainable sensitivity 
as measured by $c_{50}$ i.e., the target concentration 
resulting in $50\%$ hybridization at the spot. Regarding 
the kinetics, it yields a simple criterion for the onset 
of slowdown due to the brush formation.

Experiments using DNA chips involve many control 
parameters concerning the chip design, the sample 
preparation and the hybridization conditions. These
are outlined in Design of Oligonucleotide Microarray
Experiments together with a discussion of the resulting 
hybridization regimes and the choice of parameters 
used in our numerical calculations. Our analysis 
incorporates ingredients from the theory of polymer 
brushes. These are summarized in Background on Polymer
Brushes. This section describes the Flory version of 
the Alexander model of brushes as applied to terminally 
anchored polyelectrolytes in aquaous solution of high 
ionic strength. The model is modified to incorporate 
the effect of an impenetrable grafting surface. This 
is important in order to ensure crossover to the 
mushroom regime, of non-overlapping tails, and to 
enable comparison of the hybridization constants at 
the surface and in the bulk. Since the hybridization 
site is typically situated within the target, each 
hybridized probe carries two unhybridized tails. The 
necessary modifications are also discussed. When brush 
formation is possible, the hybridized targets also 
interact with neighboring probes. The resulting free 
energy penalty, within the Flory approximation, is 
described in Target-Probe Interactions. The free 
energies associated with the brush and with the 
target-probe interactions enable us to obtain the 
equilibrium hybridization isotherms. The derivation 
is discussed in Brush Effects--Thermodynamics of 
Hybridization. The hybridization isotherms allow to
quantify the sensitivity in terms of the corresponding 
$c_{50}$. In turn, these yield design guidelines 
relating the sensitivity to $n$, $N$ and $\Sigma_{0}$. 
Assuming, and later checking, that the hybridization 
rate at the surface is reaction controlled enables 
us to specify the hybridization and denaturation rate 
constants in the different regimes. The necessary 
background, on the hybridization kinetics in the bulk 
and the desorption dynamics out of a brush, as well as 
the resulting rate equations are discussed in Brush 
Effects--Kinetics of Hybridization. The second virial 
coefficient, $v$, specifying the interactions between 
the charged monomers of polyelectrolytes in the high 
salt regime is discussed in Appendix A. Using this $v$, 
we recover our earlier results of the $n = N$ case and 
discuss the comparison to the $N \gg n$ scenario.
In Appendix B we present an alternative derivation  
of our result for the hybridization contant in the
low surface density regime. This utilizes exact 
results thus avoiding the approximations inherent 
to the ``Alexander-Flory'' approximation.

\section*{DESIGN  OF OLIGONUCLEOTIDE MICROARRAY EXPERIMENTS}

Oligonucleotide chip experiments vary widely in 
their design. A brief summary of the possible 
designs is necessary in order to delineate the 
range of applicability of our analysis and the 
different possible regimes. To this end it is 
helpful to distinguish between three groups of 
design parameters: The chip design, the sample 
characteristics and the hybridization conditions. 
The primary parameters in the chip design (Pirrung, 
2002) are the area per probe, $\Sigma_{0}$, and 
the number of bases in the probe, $n$. $n$ values 
in the range $10$ to $30$ are typical. In this 
context one should discriminate two approaches 
to the production of oligonucleotide chips. 
In one, the probes are synthesized in situ using 
photolithography. In the other, pre-synthesized 
oligonucleotides with functionalized end groups 
are delivered to the spot. In the first approach 
it is necessary to allow for the production of 
incomplete sequences leading to polydispersity in 
$n$ (Forman et al., 1998). The reported probe 
densities within spots vary between $1.2 \times 
10^{10}$ and $4 \times 10^{13}$ probes per $cm^{2}$ 
corresponding to $2.5 \times 10^{2} \mathring{A}^{2} 
\leq \Sigma_{0} \leq 8.3 \times 10^{5} \mathring{A}^2$. 
The chip characteristics also include the nature 
of the surface treatment used to minimize non-specific 
adsorption and of the spacer chains joining the probe 
to the anchoring functionality (length, charge, 
hydrophobicity, etc.).

A key qualitative characteristic of the sample is the 
chemical nature of the targets (Graves, 1999; Lockhart 
and Winzeler, 2000; Heller, 2002). To begin, it is 
necessary to distinguish between DNA and RNA targets
which differ in two respects: First, single stranded 
(ss) RNA exhibits pronounced secondary structure (loops, 
hairpins, etc.) which is largely absent in ssDNA. Second, 
the hybridization free energy of RNA-DNA complexes is 
higher than that of DNA-DNA ones. For DNA samples, it is 
further necessary to distinguish between samples of 
double stranded (ds) DNA, as obtained from symmetric 
PCR amplification, and ssDNA samples as obtained, for 
example, using Lambda exonuclease digestion. The 
hybridization isotherms of the two types of samples are 
different (Halperin et al., 2004a). The labelling of the 
targets can also affect the hybridization behavior (Naef 
and Magnasco, 2003). Our discussion concerns samples of 
ssDNA targets assuming ideal labels that do not interfere 
with the hybridization. It focuses on the role of two 
quantitative characteristics of the sample: the number of 
bases in the target, $N$, and the molar concentration of 
the target, $c_{t}$. $N$ is determined by the choice of
primers used for the PCR amplification or by the fragmentation 
step in the sample preparation. Note that the products of 
the PCR are monodisperse while the fragmentation introduces 
polydispersity in the size of the targets. In this last 
case it is only possible to control the average size of 
the targets. Typical reported values for PCR products 
vary in the range $100 \leq N \leq 350$. The average $N$ 
resulting from the fragmentation
procedure is not always specified but the range $50 \leq N
\leq 200$ is representative. It is useful to note another 
distinction between the two procedures. Targets produced by 
PCR often have the hybridization site situated roughly in 
the middle of the target. In the case of fragmented targets, 
the location of the hybridization site is no longer controlled. 
With regard to $c_{t}$ it is helpful to stress the distinction 
between bioanalytic experiments, utilizing DNA chips to 
interrogate biological samples (See for example: Prix et al.,
2003), and physical chemistry experiments aiming to understand 
the function of DNA chips (See for example: Peterson et al., 
2002). In biology experiments $c_{t}$ is a priori unknown 
since it is set by the biological sample and its treatment. 
In marked contrast, in physical chemistry experiments the 
target concentration is imposed by the experimentalist as is 
the composition of the sample. In such experiments the target 
used is often identical in length to the probes, $n = N$. 
As noted earlier, our analysis is motivated by bioanalytical
experiments where $N \gg n$.

The hybridization conditions include the composition of 
the hybridization solution, the hybridization temperature, 
$T$, and the hybridization time, $t_{h}$. Typical 
hybridization temperatures vary over the range $30^{\circ}C 
\leq T \leq 60^{\circ}C$ depending on $n$ and the GC fraction. 
The hybridization times also vary widely with typical 
values in the range of $2h \leq t_{h} \leq 16h$. In most 
cases the hybridization solution contains $1M$ of $NaCl$.

Different hybridization regimes are possible, depending 
on the values of $n$, $N$ and $\Sigma_{0}$. To distinguish 
these regimes, it is necessary to first specify the molecular 
length scales of ssDNA and dsDNA. These are well established 
for dsDNA (Cantor and Schimmel, 1980). In the range of 
parameters considered, dsDNA is a rod-like molecule with 
each base pair contributing $3.4 \mathring{A}$ to its length. 
The radius of dsDNA is $9.5 \mathring{A}$ and its cross 
section area is $284 \mathring{A}^{2}$. We will limit our 
analysis to $\Sigma_{0} > 284 \mathring{A}^{2}$ in order 
to avoid discussion of steric hindrance to hybridization. 
The corresponding characteristics of ssDNA are not well 
established. A typical value of the monomer size is $a = 
6 \mathring{A}$ (Smith et al., 1996; Strick et al., 2003). 
The cited values of the persistence length, $l_{p}$, vary 
between $l_{p} = 7.5 \mathring{A}$ and $l_{p} = 35 
\mathring{A}$ (Mills et al., 1999). ssDNA is often 
described as a random coil though long range interactions 
are expected to give rise to swollen configurations (Turner, 
2000). In the following we will consider ssDNA as a swollen 
coil characterized by its Flory radius (Rubinstein and Colby, 
2003). This choice is dictated by our treatment of the brush, 
where the Flory radius emerges as a natural length scale. 
Accordingly, an isolated unhybridized probe occupies a 
hemisphere of radius $r_{F} \sim n^{3/5} a$ while a 
terminally hybridized target occupies a hemisphere of 
radius $R_{F} \sim (N-n)^{3/5} a \simeq N^{3/5} a$. 
As we shall discuss, the unhybridized probes do not 
interact when $r_{F}^{2} < \Sigma_{0}$. Similarly, when 
$R_{F}^{2} < \Sigma_{0}$ there is no brush regime. 
It is thus possible to distinguish between three different 
scenarios. A Langmuir regime is expected when $\Sigma_{0} 
> R_{F}^{2} > r_{F}^{2}$. Brush effects, with 
no interactions between the probes, will occur when 
$r_{F}^{2} < \Sigma_{0} <R_{F}^{2}$. Finally, when 
$\Sigma_{0} < r_{F}^{2} < R_{F}^{2}$ both the brush 
effect and probe-probe interactions play a role. All 
three scenarios occur in the reported variety of DNA chips. 

In the following we consider the role of $n$, $N$ and
$\Sigma_{0}$ in bioanalytical experiments. For brevity 
we focus on the simplest among the experimentally 
realistic situations. Thus, we consider monodispersed 
ssDNA targets and monodispersed oligonucleotide probes. 
This avoids complication due to unspecified polydispersity 
and to competitive bulk hybridization. It is convenient 
to concentrate on the $r_{F}^{2} < \Sigma_{0} < R_{F}^{2}$ 
case with $N \gg n$. As we shall see, this makes for a 
simpler discussion of the brush effects. It also allows 
us to ignore small corrections due to probe-probe 
interactions. Finally, our analysis assumes DNA chips 
with a passivated surface and probes anchored to the 
surface via short, flexible spacer chains. 

Our analysis is concerned with the modifications of the 
hybridization isotherm and rate equations as $\Sigma_{0}$ 
decreases from the Langmuir range, $\Sigma_{0} > R_{F}^{2} 
> r_{F}^{2}$, into the brush regime, $r_{F}^{2} < 
\Sigma_{0} < R_{F}^{2}$. To implement this program, it is 
helpful to identify a reference state. In the following 
we utilize a probe layer that approaches the bulk values 
for the hybridization rate and equilibrium constants. 
We argue that this is the case when the following conditions 
are satisfied. First, the surface is perfectly non-adsorbing 
to both ss and ds DNA. 
Under these conditions adsorbed states are not involved in 
the hybridization reaction and the two state approximation 
for the hybridization reaction is justified. Second, the 
probes are attached to the surface via {\it long}, flexible 
and neutral spacers. We argue that the effect of the surface 
diminishes as the length of the spacers increases. Note that 
the spacers modify two effects. One is the steric hindrance 
that occurs when the probes are directly attached to the 
surface. The other is the reduction in the number of 
accessible configurations in the vicinity of an impenetrable 
planar surface. Idealy, the reference state involves spacer 
chains that do not interact with either the probes and the 
targets. The third condition is that the distance between 
the anchored probes ensures zero probe-probe interaction 
energy, irrespective of their hybridization state. For this 
reference state, the equilibrium hybridization constant at the 
surface $K_{pt}$ approaches $K_{\overline{p}t}$, the equilibrium
hybridization constant for the bulk reaction between the
free chains. Accordingly, the hybridization isotherm in 
the small spot limit, when the hybridization at the surface 
has a negligible effect on initial molar concentration of 
the target $c_{t}$, is
\begin{equation}
\label{II1}
\frac{x_{eq}}{1-x_{eq}} = K_{pt} c_{t}.
\end{equation}
It is important to distinguish between $K_{pt}$ and 
\begin{equation}
K_{pt}^0 = \exp \left(-\frac{\Delta G_{pt}^{0}}{RT}\right),
\end{equation}
where $\Delta G_{pt}^{0}$ is the molar standard 
hybridization free energy as obtained from the nearest 
neighbor model (SantaLucia and Hicks, 2004), $T$ is 
the temperature and $R = 1.987 \, cal.mol^{-1}.K^{-1}$ 
is the gas constant. First, $K_{pt}^0$ and $\Delta 
G_{pt}^{0}$ as calculated from the nearest neighbor 
model are identical for all $N \geq n+2$. They allow, 
at most, for the effect of two dangling ends. Second, 
this model incorporates only nearest neighbor interactions 
along the backbone of the chain. It thus assumes that the 
oligonucleotide adopts the configuration of an ideal random 
coil. In particular, $\Delta G_{pt}^0$ does not account for
excluded volume interactions between the monomers. In 
addition, $\Delta G_{pt}^{0}$ clearly does not allow for 
the effect of the impenetrable wall or for the interactions 
between the hybridized targets or between them and the 
neighboring probes. These additional terms and their 
effect on the hybridization isotherm will be discussed 
in the following three sections.

Our choice of the parameters used in the numerical 
calculations is based on two experimental systems. 
One, of Guo et al. (1994), utilized probes of length 
$n = 15$ with PCR produced targets of length $N = 157$ 
or $347$. Both ssDNA and dsDNA were investigated, 
the area per probe was varied in the range $300 
\mathring{A}^{2} \leq \Sigma_{0} \leq 3000 
\mathring{A}^{2}$ and the hybridization was carried 
out at $T = 30^{\circ}C$. The hybridization times 
varied with $N$ being $t_{h} = 2-3h$ for $N = 157$ 
and $t_{h} = 6-8h$ for $N=347$. Note that in this 
study some of the data corresponds to the $\Sigma_{0} 
< r_{F}^{2} < R_{F}^{2}$ regime where probe-probe 
interactions are not negligible. The second system 
is the Affymetrix GeneChip E. Coli Antisense Genome 
Array (Affymetrix, 2004). In this case, probes of 
length $n = 25$ hybridize with fragmented, thus 
polydispersed, ds cDNA targets with average length 
in the range $50 \leq N \leq 200$. The hybridization 
is carried out at $T = 45^{\circ}C$ for $t_{h} = 16h$. 
A rough approximation of $\Sigma_{0}$ for Affymetrix 
chips was obtained from the estimated density of 
functional groups in the substrate prior to the 
synthesis of the probes: $27 \, pM/cm^{2}$, and the 
step-wise yield of the synthesis, $\sim 90\%$. Only 
$14 \, pM/cm^{2}$ attain $n \geq 6$ (Forman et al., 
1998). This estimate yields $\Sigma_{0} \geq 1200 
\mathring{A}^{2}$. In both systems the hybridization 
was carried out in a solution containing $1M$ of 
$NaCl$. The base sequence of the probes considered 
in the calculations and their thermodynamic parameters 
for hybridization, as calculated using the nearest 
neighbor model with a perfectly matched target 
(SantaLucia, 1998; Peyret et al., 1999; HyTher$^{TM}$), 
are specified in Table 1.

Table 1: The thermodynamic parameters utilized 
in the numerical calculations correspond to two 
probes: (i) The $n = 15$ wild type probe $p1$  
($5'-CGT CCT CTT CAA GAA-3'$) incorporates the 
codon 406 of exon 4 of the human tyrosinase 
gene. The $N = 157$ and $347$ targets incorporate 
the perfect complementary segment $5'-TTC TTG AAG 
AGG ACG-3'$ (Guo et al., 1994). (ii) The Affymetrix 
$E. Coli$ Antisense $n = 25$ probe $p2$ annotated 
AFFX-BioB-5\_at:242:77, with interrogation point 
$177$, corresponds to the sequence $5'-AGA TTG 
CAA ATA CTG CCC GCA AAC G-3'$. The fragmented 
cDNA targets incorporate the perfect complementary 
sequence $5'-CGT TTG CGG GCA GTA TTT GCA ATC T-3'$. 
The parameters are calculated from the nearest 
neighbor model (SantaLucia, 1998; Peyret et al., 
1999; HyTher$^{TM}$) using the HyTher$^{TM}$ program 
with a $1M$ $NaCl$ salt concentration. Since the 
targets are longer than the probes two dangling 
ends are invoked.

\begin{center}
\begin{tabular}[c]{ccccc}
probe \ & $\Delta H_{pt}^{0}$ & $\Delta S_{pt}^{0}$ &
\ $\Delta G_{pt}^{0} (30^{\circ}C)$ \ &
\ $\Delta G_{pt}^{0} (45^{\circ}C)$ \ \\
& \ $kcal.mol^{-1}$ \ & \ $cal.mol^{-1}.K^{-1}$ \ & 
$kcal.mol^{-1}$ & $kcal.mol^{-1}$ \\
$p1$ & -121.00 & -334.06 & -19.73 & -14.72\\
$p2$ & -203.30 & -546.32 & -37.69 & -29.49
\end{tabular}
\end{center}

\section*{BACKGROUND ON POLYMER BRUSHES}

\begin{figure}
\caption{A schematic picture of the hybridization 
oflong targets at a layer of short probes. 
For simplicity we depict the case of targets with 
a terminal hybridization site, when each hybridized 
probe carries a long ssDNA tail. Three regimes occur:
a) In the $1:1$ regime the distance between the probes, 
$\Sigma_0^{1/2}$, is large and each hybridized target 
can only interact with its own probe. There is no 
crowding of the tails. b) In the $1:q$ regime the probe 
density is higher. At low hybridization fraction each
target interacts with $q = R_F^2/\Sigma_0$ probes. 
c) As the hybridization fraction increases the 
hybridized targets begin to crowd each other thus
forming a brush with an area per chain $R_F^2 > 
\Sigma > \Sigma_0$. Note that in the general case the 
hybridization site is situated roughly in the middle 
of the target and each hybridized probe carries two 
tails (d).}
\end{figure}

Polymer brushes are formed by chains with one monomer 
anchored to a planar surface (Milner, 1991; Halperin 
et al., 1992). In the simplest case, the anchoring
moiety is the terminal monomer. When the area per 
chain, $\Sigma$, is large the chains do not crowd 
each other. In this ``mushroom'' regime, the chains 
may be roughly considered as occupying hemispheres 
whose radius is comparable to the Flory radius of 
the free chain, $R_{F}$. When the surface density
increases such that $\Sigma \leq R_{F}^{2}$, the 
chains begin to crowd each other thus forming a 
``brush''. In the brush regime the chains stretch 
out along the normal to the surface so as to decrease 
the monomer concentration, $c$, and the number of 
repulsive monomer-monomer contacts. A simple description 
that captures the leading behavior of brushes is 
provided by the Alexander model (Alexander, 1977; 
Milner, 1991; Halperin et al., 1992). Within it 
the concentration profile of the brush is modeled by 
a step function of height $H$ i.e., $c = N/H\Sigma$ 
at altitudes up to $H$ above the surface and $c = 0$ 
for higher altitudes. All the free ends are assumed 
to straddle the outer boundary of the brush at height 
$H$. In the following we will use the Flory-style
version of the model, ignoring scaling corrections. 
The regime of validity of this meanfield approach 
for semiflexible chains is expanded in comparison
to that of flexible polymers (Birshtein and Zhulina, 
1984). This justifies the use of the ``Alexander-Flory''
model where the free energy per chain in a brush is
\begin{equation}
\label{III1}
\frac{G}{kT} = v \frac{N^{2}}{\Sigma H} + 
\frac{H^{2}}{N a l_{p}} - \ln \frac{H}{N a}
\end{equation}
where $k$ is the Boltzmann constant.
The first term allows for the monomer-monomer 
interactions. It is of the form $v c^{2} V_{chain}$ 
where $v$ is the second virial coefficient and
$V_{chain} = \Sigma H$ is the volume per chain. 
The second accounts for the entropy loss incurred 
because of the stretching of a Gaussian chain,
comprising of $N a/l_{p}$ persistent sequences of 
length $l_{p}$, along the normal to the surface. 
Here $a$ is the monomer size, $l_{p}$ is the
persistence length of the chain and the span of 
the Gaussian unswollen coil is $R_{0} = \sqrt{Nal_{p}}$. 
The last term arises because the impenetrable surface
carrying the anchoring site reduces the number of 
accessible configurations of the tethered chain. 
For a Gaussian chain with a free end at altitude $H$ 
the number is reduced by a factor of $H l_{p}/R_{0}^{2}$ 
(DiMarzio, 1965). This contribution is often ignored 
because it has a negligible effect on the equilibrium 
dimensions of the chains. It leads however to a 
significant modification of the hybridization constant 
at the surface. The last two terms of Eq.\ref{III1} 
apply, in this form, when $N a \gg l_{p}$. We have
omitted a term allowing for the entropy associated
with the placement of the free end. This is because
the Alexander model assumes that all free ends are 
constrained to the brush boundary. For simplicity we
ignore, here and in the following, numerical factors 
of order unity. Minimization of $G$ with respect to 
$H$ yields the equilibrium values of $G_{brush}$ and $H$
\begin{eqnarray}
\frac{G_{brush}}{kT} & = & N \left(\frac{a^{2}}{\Sigma}
\right)^{2/3} \left(\frac{a}{l_p}\right)^{1/3} \left(
\frac{v}{a^3}\right)^{2/3} - \ln \left[ \left(
\frac{a^2}{\Sigma}\right)^{1/3} \left(\frac{l_p}{a}
\right)^{1/3} \left(\frac{v}{a^3}\right)^{1/3} \right],
\label{III2}\\
\frac{H}{a} & = & N \left(\frac{a^2}{\Sigma}\right)^{1/3} 
\left(\frac{l_{p}}{a}\right)^{1/3} \left( \frac{v}{a^{3}}
\right)^{1/3}.
\end{eqnarray}
In the mushroom regime, the chains occupy a hemisphere 
of radius 
\begin{equation}
\label{EqRF}
\frac{R_{F}}{a} = N^{3/5} \left(\frac{l_{p}}{a}
\right)^{1/5} \left( \frac{v}{a^{3}} \right)^{1/5}.
\end{equation}
Accordingly, the free energy per chain in the mushroom 
regime, $G_{mush}$, is set by the requirement $G_{mush} 
= G_{brush}$ at the mushroom-brush boundary when $\Sigma 
= R_{F}^{2}$ and $H = R_F$ thus leading to
\begin{equation}
\frac{G_{mush}}{kT} = N^{1/5} \left(\frac{a}{l_{p}} 
\right)^{3/5} \left(\frac{v}{a^{3}}\right)^{2/5} - 
\ln \left[N^{-2/5} \left( \frac{l_p}{a}\right)^{1/5} 
\left(\frac{v}{a^{3}}\right)^{1/5} \right]. \label{III3}
\end{equation}
As noted earlier, the properties of the chains in the 
mushroom regime are comparable to those of free coils. 
In turn, the free coil behavior is specified by the 
free energy (De Gennes, 1979)
\begin{equation}
\frac{G}{kT} = v \frac{N^2}{r^3} + \frac{r^2}{N a l_p}
\label{IIIa}
\end{equation}
leading, upon minimization with respect to the radius 
$r$, to $R_F$ as given by Eq.\ref{EqRF} and to the 
equilibrium free energy of a coil
\begin{equation}
\frac{G_{coil}}{kT} = N^{1/5} \left(\frac{a}{l_{p}} 
\right)^{3/5} \left(\frac{v}{a^{3}}\right)^{2/5}.
\label{IIIb}
\end{equation}
The difference between $G_{mush}$ and $G_{coil}$ 
is due to the logarithmic correction $-\ln (R_F/Na)$ 
arising from the wall effect. 

Within the approach described above, the nature of 
the grafted chain is specified by three parameters, 
the monomer size $a$, the persistence length $l_{p}$, 
and the second virial coefficient associated with 
monomer-monomer interactions $v$. For the case of a 
brush formed by polyelectrolyte chains in aqueous 
solution of high ionic strength, ``high salt'', $v$ 
can be approximated by (Pincus, 1991; Appendix A)
\begin{equation}
v = \frac{2 \pi}{3} a^{3} \left(1-\frac{\theta}{T}
\right) + 2 \pi l_{B} r_{D}^{2}.
\label{III4}
\end{equation}
The first term allows for the hard core repulsion 
between the monomers and for a weak, long ranged, 
van der Waals attraction between them. Here $\theta$ 
is the theta temperature where $v$ of a neutral 
chain vanishes thus leading to the behavior of 
an ideal Gaussian coil. This term by itself is 
used to describe the behavior of neutral polymers 
(Rubinstein and Colby, 2003). The second term 
arises from the screened electrostatic interactions 
between the singly charged monomers. Here $l_{B} =
e^{2}/\epsilon k T$ is the Bjerrum length
(Evans and Wennerstr\"{o}m, 1994) where $\epsilon$ 
is the dielectric constant, $k$ the Boltzmann 
constant and $T$ the temperature. In water, 
with $\epsilon \simeq 80$ at room temperature, 
$l_{B} \simeq 7 \mathring{A}$. Note that the
variation of $\epsilon$ with $T$ contributes to 
the $T$ dependence of $l_{B}$. The Debye length 
$r_D$ characterizes the range of the screened 
electrostatic interactions in a salt solution 
(Evans and Wennerstr\"{o}m 1994). For a $1:1$
salt with number concentration of ions $c_{s}$, 
$r_{D} = 1/\sqrt{8\pi l_{B} c_{s}}$ thus, in a 
$1M$ solution $r_{D} = 3 \mathring{A}$. In our 
model, the presence of the $2 \pi l_{B} r_{D}^{2}$ 
term in $v$ distinguishes polyelectrolyte brushes
from neutral ones. It is important to stress the 
limitations of approximating $v$ by Eq.\ref{III4}.
It corresponds to the interaction between individual 
charged spherical monomers. For cylindrical 
non-charged monomers $v \simeq l_p^2 \, a$ rather 
than $v \simeq a^3$ (Rubinstein and Colby, 2003). 
Furthermore, this description does not allow for the 
contribution of Hydrogen bonds with water nor for the 
effect of correlations on the electrostatic interactions. 
Finally, the appropriate $\theta$ temperature 
remains to be determined. With these caveats in mind, 
the second term is roughly comparable to $2 \pi a^3/3$ 
and should be dominant for $T \gtrsim \theta$. 
As a result $v$ is comparable to $2 \pi a^{3}/3$ and 
the swelling behavior of the chain is similar to that 
of a neutral chain in an athermal solvent (De Gennes, 
1979). In other words, even short chains swell to their 
Flory radius. We should add that by using $v \simeq 2 
\pi l_B r_D^2$ we are able to recover our earlier 
results (Halperin et al., 2004a) for the case of 
$n = N$ (Appendix A).

In the Flory type approach, described above, the 
equilibrium state is determined by a global balance 
of the osmotic pressure of the monomers and the
restoring elastic force of the stretched Gaussian 
chains. A more refined analysis of the brushes, 
utilizing self consistent field (SCF) theory, is
possible. This avoids the assumptions of uniform 
stretching and step-like concentration profiles. 
It yields the same functional forms for the
characteristic height, $H$, and for $G_{brush}$ 
but with somewhat different numerical prefactors. 
With these reservations in mind we utilize the 
simplest approach, described earlier, because it 
typically yields the correct leading behavior in 
similar systems. A SCF theory is necessary for the 
description of effects that depend strongly on the 
details of the concentration profile and the 
distribution of the free ends.

Our discussion thus far concerned brushes anchored 
to the surface by the terminal head group. In DNA 
chips the situation is often different in that the
hybridization site, the anchoring functionality, 
is located roughly at the middle of the target. 
As a result, each hybridized probe carries two 
unhybridized tails (Fig.1d) of length $N_{1}$ and 
$N_{2} = N_{1}(1+\alpha)$ such that $N_{1} + N_{2} 
+ n = N$. In considering the effect of this feature 
note that, in the brush regime, the details of the 
anchoring functionality are screened with a distance 
$\Sigma^{1/2}$ from the surface. As a result, 
it is possible to estimate the modification of 
$G_{brush}$ and $H$ in two cases, $N_{1} = N_{2} 
\gg n$ and $N_{2} \gg N_{1} \gg n$. When $N_{1}
= N_{2}$ the resulting brush is similar to that 
formed by chains of length $N/2$ but with an area 
per chain $\Sigma/2$. In this case $G_{brush}$ is 
larger by a factor $2^{2/3} \simeq 1.6$ while 
$H$ is smaller by a factor $2^{2/3}$ in comparison 
to the values found for a brush of terminally 
anchored chains of length $N$ and area per chain 
$\Sigma$. In the limit of $N_{2} \gg N_{1} \gg n$ 
the resulting brush may be considered as bidispersed, 
comprising an equal number of chains of length 
$N_{1}$ and $N_{2}$. Such a bidispersed brush can 
be described as a superposition of two brushes 
(Birshtein et al., 1990). A simple two layer
model incorporates an inner brush of chains of 
length $N_{1}$ and area per chain of $\Sigma/2$
and an outer brush formed by chains of length 
$N_{2} - N_{1} = \alpha N_1$ and area per chain 
$\Sigma$ at the distal boundary of the inner brush. 
Within the Flory approximation this scheme leads 
to $\widetilde{G}_{brush} = \frac{\alpha + 
2^{5/3}}{\alpha + 2} G_{brush}$ and $\widetilde{H} 
= \frac{\alpha + 2^{1/3}}{\alpha + 2} H$ where 
$G_{brush}$ and $H$ correspond to a 
monodispersed brush of chains of length $N$ with 
an area per chain $\Sigma$. Note that $\alpha = 0$ 
corresponds to $N_1 = N_2$ while $\alpha \gg 1$ to 
$N_2 \gg N_1$. In both cases, the effect is to 
modify $G_{brush}$ and $H$, as obtained earlier by 
a multiplicative factor of order unity. In keeping 
with our policy we will omit these numerical 
factors in the interest of simplicity.

\section*{TARGET-PROBE INTERACTIONS}

The preceding discussion of brushes allows
for the interactions among the hybridized 
targets and the effects of the impenetrable 
wall. However, the brush regime is only attainable 
when the hybridized targets can interact with 
neighboring probes, thus giving rise to an 
additional contribution to the free energy of the 
system. In discussing the target-probe interactions 
it is useful to distinguish between three regimes. 
When $\Sigma_{0} > R_{F}^{2} > r_{F}^{2}$ the 
hybridized targets can not crowd each other. 
Roughly speeking, each one may be considered to 
occupy a hemisphere of radius $R_{F}$ containing a 
single probe that is hybridized to the target 
(Fig.1a). Since each target interacts with a single 
probe we will refer to this regime as $1:1$. 
Our principle interest is in the two regimes
that occur when $R_{F}^{2} > \Sigma_{0} > r_{F}^{2}$. 
When the hybridization degree $x$ is sufficiently 
small each target will occupy, as before, a hemisphere 
of radius $R_{F}$. However, it will now interact with 
$q = R_{F}^{2}/\Sigma_{0}$ probes (Fig.1b). We will 
thus refer to this regime as $1:q$. Note that in the 
polymer science nomenclature both $1:1$ and $1:q$ 
regimes fall into the ``mushroom'' range, when the 
tethered chains do not overlap. The brush threshold 
occurs at $x = x_{B}$ when the hemispheres occupied 
by the different targets come into grazing contact. 
For a surface of total area $A_{T}$ the area per 
hybridized target is $\Sigma = A_{T}/x N_{T} = 
\Sigma_{0}/x$ where $N_{T}$ is the total number 
of probes. $x_{B}$ corresponds to $\Sigma = R_F^2$ 
or $x_{B} = \Sigma_{0}/R_{F}^{2} = 1/q$. When $x$ 
exceeds $x_{B}$ the hybridized targets begin to 
overlap thus forming a brush (Fig.1c). Since the 
area per chain in this regime decreases as $\Sigma 
= \Sigma_{0}/x$ the target experiences interactions 
only with $x^{-1} < q$ probes.

To estimate the free energy of interactions 
between the target and the probes, in the 
spirit of the Flory approach, we assume that 
each probe contributes an interaction free 
energy $G_{int}/kT = vnc$. Here $c$ is the 
monomer concentration within the monomer cloud 
formed by the hybridized targets i.e., we assume 
the interaction with the probes does not affect 
$c$ as obtained in our earlier discussion of the 
mushroom and brush regimes. As we shall elaborate 
later, this assumption is justified only when 
$G_{int} \ll G_{coil}(N)$ or
\begin{equation}
\Sigma_{0} \gg N^{1/5} n a^2 \left( \frac{l_p}{a} 
\right)^{2/5} \left( \frac{v}{a^{3}} \right)^{2/5}.
\label{condition}
\end{equation}
In the $1:1$ regime each hybridized target 
occupies a hemisphere of radius $R_{F}$ 
incorporating a single probe. Accordingly 
$G_{int}^{1:1}/kT = vnc$ with $c = N/R_{F}^{3}$ 
thus leading to
\begin{equation}
\frac{G_{int}^{1:1}}{kT} = \frac{n}{N^{4/5}} \left(
\frac{a}{l_{p}}\right)^{3/5} \left(\frac{v}{a^{3}}
\right)^{2/5}.
\label{T1}
\end{equation}
This estimate is reasonable when $N \gg n$ such 
that the region occupied by the unhybridized target 
is sufficiently large so as to encompass the 
hybridized probe. Roughly speaking, this implies
$(N-n)^{3/5} a > 3.4 \, n \, \mathring{A}$. 
Within the $1:q$ regime each hybridized target 
interacts with $q = R_{F}^{2}/\Sigma_{0}$ probes. 
Accordingly $G_{int}^{1:q}/kT = v (R_{F}^{2}/
\Sigma_{0})nc$ with $c=N/R_{F}^{3}$ or
\begin{equation}
\frac{G_{int}^{1:q}}{kT} = N^{2/5} n \left(
\frac{a^{2}}{\Sigma_{0}}\right) \left(\frac{a}{l_{p}}
\right)^{1/5} \left(\frac{v}{a^{3}}\right)^{4/5}.
\label{T2}
\end{equation}
$G_{int}^{1:1}$ and $G_{int}^{1:q}$ are independent 
of $x$. In marked contrast $G_{int}^{B}$, accounting 
for the target-probe interactions in the brush regime, 
varies with $x$. This variation arises because of the 
$x$ dependence of the monomer concentration within 
the brush, $c_{brush} = N/\Sigma H$ where $\Sigma 
\sim 1/x$ and $H \sim x^{1/3}$. $G_{brush}(x)$ is 
obtained from Eq.\ref{III2} upon replacing $\Sigma$ 
by $\Sigma_{0}/x$. Within the Flory approach the 
total interaction free energy between the targets and 
the probes is $v N_{T} n c_{brush}$. The interaction 
free energy per hybridized target is thus $v n 
c_{brush}/x$ or
\begin{equation}
\frac{G_{int}^{B}}{kT} = \frac{v n N}{\Sigma_{0}H} =
nx^{-1/3} \left(\frac{a^{2}}{\Sigma_{0}}\right)^{2/3}
\left(\frac{a}{l_{p}}\right)^{1/3}
\left(\frac{v}{a^{3}}\right)^{2/3}.
\label{T3}
\end{equation}

The condition Eq.\ref{condition} ensures that the
interaction term $G_{int}^{1:q}$ is a weak perturbation 
to the Flory free energy of the mushroom $G_{mush}(N)$. 
When this requirement is not satisfied the chain span 
exceeds the Flory radius. This is an unphysical result 
since the interactions driving the extra swelling are 
confined to the surface. In this case the chain can no 
longer be assumed to occupy a hemispherical region 
encompassing the probes. The uniform monomeric 
distribution inherent to the Flory approach should 
be refined so as to reflect locally stretched 
configurations allowing to avoid the probes. 
For simplicity we will not consider this regime.

\section*{BRUSH EFFECTS--THERMODYNAMICS OF HYBRIDIZATION}

Having obtained the free energy terms associated with 
target-target and target-probe interactions at the 
surface, we are in a position to investigate their
effect on the hybridization isotherm. To simplify 
the equations we set $v = a^{3}$ and $l_{p}=a$. 
The hybridization isotherm is determined by the 
equilibrium condition of the hybridization reaction 
$p + t \rightleftharpoons pt$ at the probe layer
that is $\mu_{pt} = \mu_{p} + \mu_{t}$ where 
$\mu_{i}$ is the chemical potential of species $i$. 
Here $p$ and $t$ signify single stranded probe and 
target while $pt$ is the hybridized probe-target pair. 
We first consider $\mu_{t}$. In practice, the molar
concentration of the targets, $c_{t}$, is only weakly 
diminished by the hybridization reaction and it is 
reasonable to assume that $c_{t}$ is constant. The 
generalization to the opposite case, when this small 
spot approximation fails, is straightforward (Halperin 
et al., 2004a). Since the target solution is dilute
and the ionic strength of the solution is high, 
electrostatic interactions between the targets are 
screened. Consequently $\mu_{t}$ assumes the weak
solution form
\begin{equation}
\mu_{t} = \mu_{t}^{0} + N_{Av} G_{coil}(N) + RT 
\ln c_{t} \label{V1}
\end{equation}
where $\mu_{t}^{0}$ is the chemical potential of 
the standard state of the hybridization site and
$N_{Av}$ is the Avogadro number. We choose a 
standard state such that $\mu_{pt}^0 - \mu_p^0 
- \mu_t^0 = \Delta G_{pt}^0$ as given by the 
nearest neighbor method. As discussed earlier, 
this implies a standard state having an ideal 
coil configuration. When the hybridization site 
is within the target, it also reflects the 
contribution of two dangling ends. $G_{coil}(N)$, 
as given by Eq.\ref{III3}, allows for the swelling 
of the free coil due to excluded volume and 
electrostatic interactions. Strictly speaking, 
$\mu_{t} = \mu_{t}^{0} + N_{Av} G_{coil} + RT 
\ln a_{t}$ where $a_{t}$ is the activity (Moore, 
1972). The dimensionless $a_{t}$ is related to 
the molar concentration of targets $c_{t}$ via 
$a_{t} = \gamma c_{t}$ where $\gamma$ is the 
activity coefficient. Since $\gamma \rightarrow 
1$ as $c_{t} \rightarrow 0$ we will, for simplicity, 
express $\mu_{t}$ by Eq.\ref{V1} noting that the 
molar $c_{t}$ in this expression is dimensionless.

It is useful to first specify $K_{pt}$ of the 
reference state corresponding, as discussed in
Design of Oligonucleotide Microarray Experiments, 
to $K_{\overline{p}t}$ of the bulk reaction 
$\overline{p} + t \rightleftharpoons \overline{p} 
t$ where $\overline{p}$ denotes a free probe chain. 
To this end we need
\begin{equation}
\mu_{\overline{p}} = \mu_p^0 + N_{Av} G_{coil} (n)
+ RT \ln c_{\overline{p}}
\end{equation}
and
\begin{equation}
\mu_{\overline{p} t} = \mu_{pt}^0 + N_{Av} [G_{coil}
(N-n) + G_{1:1}] + RT \ln c_{\overline{p}t}.
\end{equation}
The equilibium condition $\mu_{t} + \mu_{\overline{p}}
= \mu_{\overline{p}t}$ yields $K_{\overline{p}t} =
K_{pt}$ with 
\begin{eqnarray}
K_{pt} & = & \exp \left\{ -\frac{\Delta G_{pt}^0 + 
N_{Av} [G_{coil}(N-n) + G_{int}^{1:1} - G_{coil}(n)
- G_{coil}(N)]}{RT} \right\} \\
\nonumber & = & K_{pt}^0 \exp \left[ n^{1/5} - 
\frac{n}{N^{4/5}} \right]
\end{eqnarray}
where $K_{pt}^0 = \exp (-\Delta G_{pt}^0/RT)$ and 
$\Delta G_{pt}^0 = \mu_{pt}^0 - \mu_p^0 - \mu_t^0$.
$K_{pt} > K_{pt}^0$ because the hybridization 
results in the formation of a rodlike ds domain 
whose monomers experience only short-range interactions 
with each other but also long-range interactions with 
the monomers of the unhybridized ss tails.

The chemical potentials $\mu_{pt}$ and $\mu_{p}$ 
are specified by the free energy per probe site 
of the surface, $\gamma_{site}$. In the $1:1$ 
regime, when $\Sigma_{0} > R_{F}^{2} > r_{F}^{2}$, 
there is no mutual interaction between the probes 
or between the targets. The molar free energy of 
probe sites is
\begin{equation}
\gamma_{site} = \gamma_{0} + x \left[ \mu_{pt}^{0}
+ N_{Av} G_{mush}(N-n) + N_{Av} G_{int}^{1:1} \right] 
+ (1-x) \left[ \mu_{p}^{0} + N_{av} G_{mush}(n) 
\right] - T S[x]. \label{VIa}
\end{equation}
Here $\gamma_{0}$ is the free energy density of the 
bare surface while $\mu_{pt}^{0}$ and $\mu_{p}^{0}$ 
denote the chemical potentials of the hybridized
and non-hybridized probes in the standard state. 
As noted before, the standard state of $p$ is an 
ideal coil with no excluded volume interactions. 
The two $G_{mush}$ terms allow for the excluded 
volume and screened electrostatic interactions as 
well as for the effect of the impenetrable wall. 
$G_{mush}(N-n)$ accounts for the monomer-monomer 
interactions of the unhybridized tail of $pt$ 
while $G_{mush}(n)$ allows for the contribution 
of the unhybridized probe. $G_{int}^{1:1}$ 
reflects the electrostatic and excluded volume 
interactions between the hybridized target and its 
own probe. The mixing entropy per mole of $p$ and $pt$ 
sites is $S[x] = - R [ x \ln x + (1-x) \ln (1-x)]$.
The equilibrium condition  $\mu_{pt}=\mu_{p}+\mu_{t}$ 
can be expressed in terms of the exchange chemical 
potential of the hybridized probe, $\mu_{pt}^{ex} = 
\mu_{pt} - \mu_{p} = \partial \gamma_{site}/\partial x$, 
as $\mu_{pt}^{ex} = \mu_{t}$. The hybridization isotherm, 
thus obtained, assumes the familiar Langmuir form
\begin{eqnarray}
\frac{x_{eq}}{1-x_{eq}} & = & c_{t} K_{pt}^{1:1} = 
c_{t} K_{pt}^0 \exp\left[-\frac{G_{mush}(N-n) + 
G_{int}^{1:1} - G_{mush}(n) - G_{coil}(N)}{kT} \right]
\label{VIb}\\
& \simeq & c_{t} K_{pt} \left(\frac{n}{N}\right)^{2/5}.
\nonumber
\end{eqnarray}
$K_{pt}^{1:1}$ is smaller than $K_{pt}$ because of 
the effect of an impenetrable wall giving rise to 
the $(n/N)^{2/5}$ factor reflecting the reduction 
in the number of configurations available to the 
unhybridized tail of $pt$. 

In the $\Sigma_{0} > R_{F}^{2} > r_{F}^{2}$ range 
the hybridization behavior is independent of $x$. 
As noted earlier, an $x$ dependence is expected 
when $R_{F}^{2} > \Sigma_{0} > r_{F}^{2}$. 
We first discuss the $1:q$ regime occurring when 
$x < x_{B}$. $\gamma_{site}$ in this range is 
similar to the one describing the $1:1$ regime.
The only difference is the replacement of 
$G_{int}^{1:1}$ by $G_{int}^{1:q}$, thus allowing 
for the interactions between a hybridized target 
and $q > 1$ probes. The hybridization isotherm 
as obtained from $\mu_{pt}^{ex} = \mu_{t}$ is
\begin{eqnarray}
\frac{x_{eq}}{1-x_{eq}} & = & c_{t} K_{pt}^{1:q} 
= c_{t} K_{pt}^0 \exp \left[-\frac{G_{mush}(N-n) 
+ G_{int}^{1:q} - G_{mush}(n) - G_{coil}(N)}{kT}
\right] \label{VI2b}\\
& \simeq & c_{t} K_{pt} \left(\frac{n}{N}\right)^{2/5}
\exp \left[ -\frac{n}{N^{4/5}}(q-1) \right]. \nonumber
\end{eqnarray}
As in the $1:1$ regime, the hybridization isotherm 
is of the Langmuir form. The equilibrium constant, 
$K_{pt}^{1:q}$ is however smaller than $K_{pt}^{1:1}$ 
because $G_{int}^{1:q}$ is larger than $G_{int}^{1:1}$ 
by a factor of $q = R_{F}^{2}$/$\Sigma_{0} = N^{6/5} 
a^2/\Sigma_0$.

When $\Sigma \leq R_{F}^{2}$ or $x \geq x_{B} = 
\Sigma_{0}/R_{F}^{2} \simeq N^{-6/5} \Sigma_{0}/a^{2}$ 
the hybridized targets begin to crowd each other
and form a brush. This crossover occurs at $x_{eq} 
= x_B$ corresponding to 
\begin{equation}
c_{B} =  \frac{\Sigma_{0}}{R_{F}^{2}-\Sigma_{0}}
\frac{1}{K_{pt}^{1:q}} = \frac{\Sigma_{0}}{R_{F}^{2}-
\Sigma_{0}}\frac{1}{K_{pt}} \left(\frac{N}{n}\right)^{2/5}
\exp \left[ \frac{n}{N^{4/5}} (q-1) \right]. 
\end{equation}
$\gamma_{site}$ of the brush regime,
\begin{equation}
\gamma_{site} = \gamma_{0} + x [ \mu_{pt}^{0} + 
N_{Av} G_{brush}(x) + N_{av} G_{int}^{B}(x) ] + 
(1-x) [ \mu_{p}^{0} + N_{Av} G_{mush}(n) ]  - T S[x],
\label{VI3}
\end{equation}
is distinctive in two respects. First, $G_{mush}(N-n)$ 
is replaced by an $x$ dependent free energy of a chain 
in a brush, $G_{brush}(x)$. Second, the term allowing 
for the target-probe interactions, $G_{int}^{B}(x)$, 
is also a function of $x$. The hybridization isotherm, 
obtained as before, is
\begin{eqnarray}
\frac{x_{eq}}{1-x_{eq}} &  = & c_{t} K_{pt}^{B}(x_{eq})
= c_{t} K_{pt}^0 \exp \left[ -\frac{G_{brush}(x_{eq}) + 
G_{int}^{B}(x_{eq}) - G_{mush}(n) - G_{coil}(N)}{kT} 
\right] \label{VI5}\\
& \simeq & c_{t} K_{pt} \left( \frac{n^{6/5} a^{2} 
x_{eq}}{\Sigma_{0}} \right)^{1/3} \! \exp \left\{
\frac{n}{N^{4/5}} - [ N x_{eq}^{2/3}- N x_{B}^{2/3} 
+ n x_{eq}^{-1/3} ] \left(\frac{a^{2}}{\Sigma_{0}} 
\right)^{2/3} \right\}. \nonumber
\end{eqnarray}
The $N^{1/5}$ term, arising from $G_{coil}(N)$ is
expressed as $N x_B^{2/3} (a^2/\Sigma_0)^{2/3}$ to
underline the crossover behavior at $x_B$. 
By construction, this isotherm is only meaningful 
when $c_t > c_{B}$ so that $x > x_{B}$. It deviates 
strongly from the Langmuir form because of the $x$
dependence of $G_{brush}$ and $G_{int}^{B}$.

\begin{figure}
\includegraphics[width=10cm]{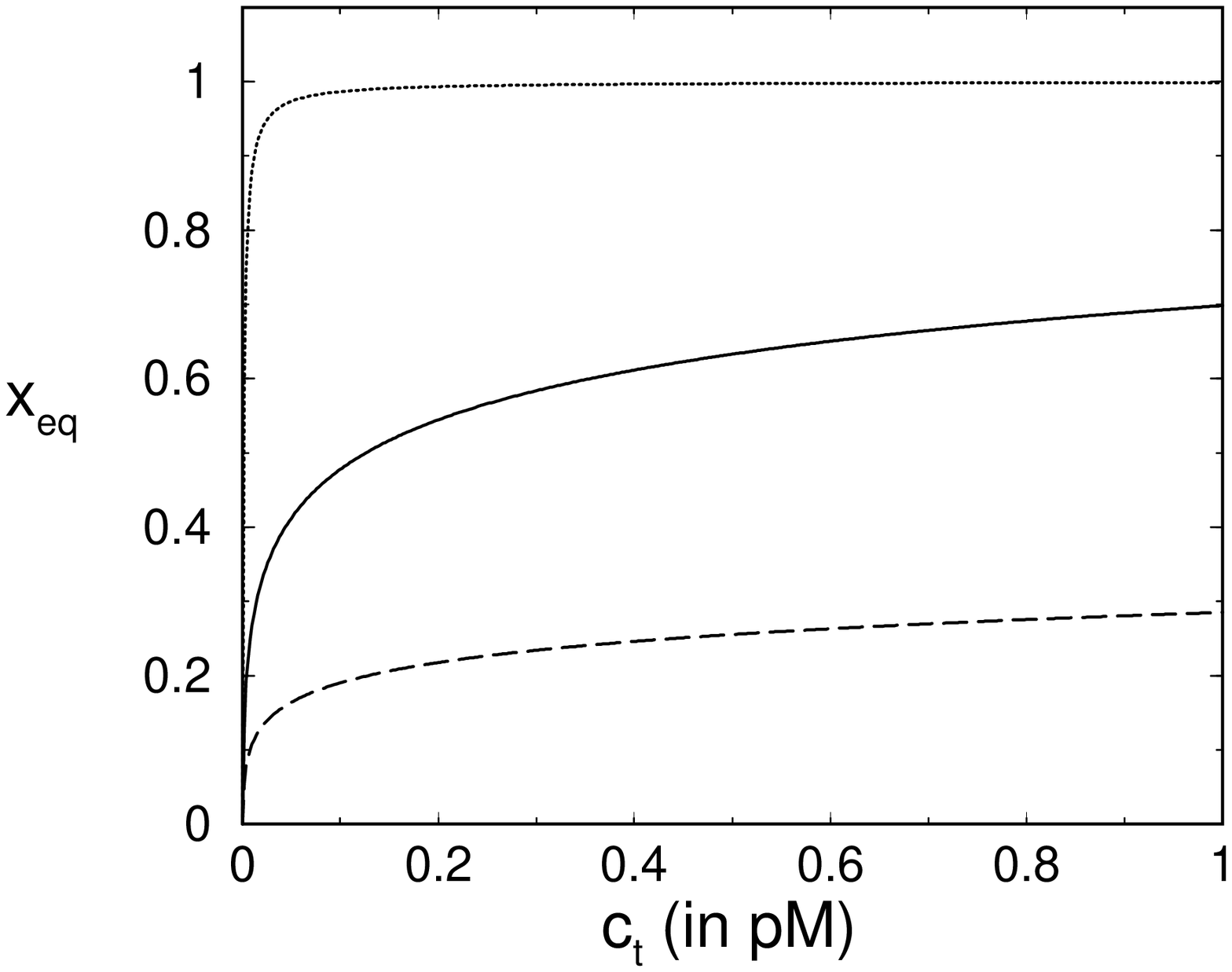}
\includegraphics[width=10cm]{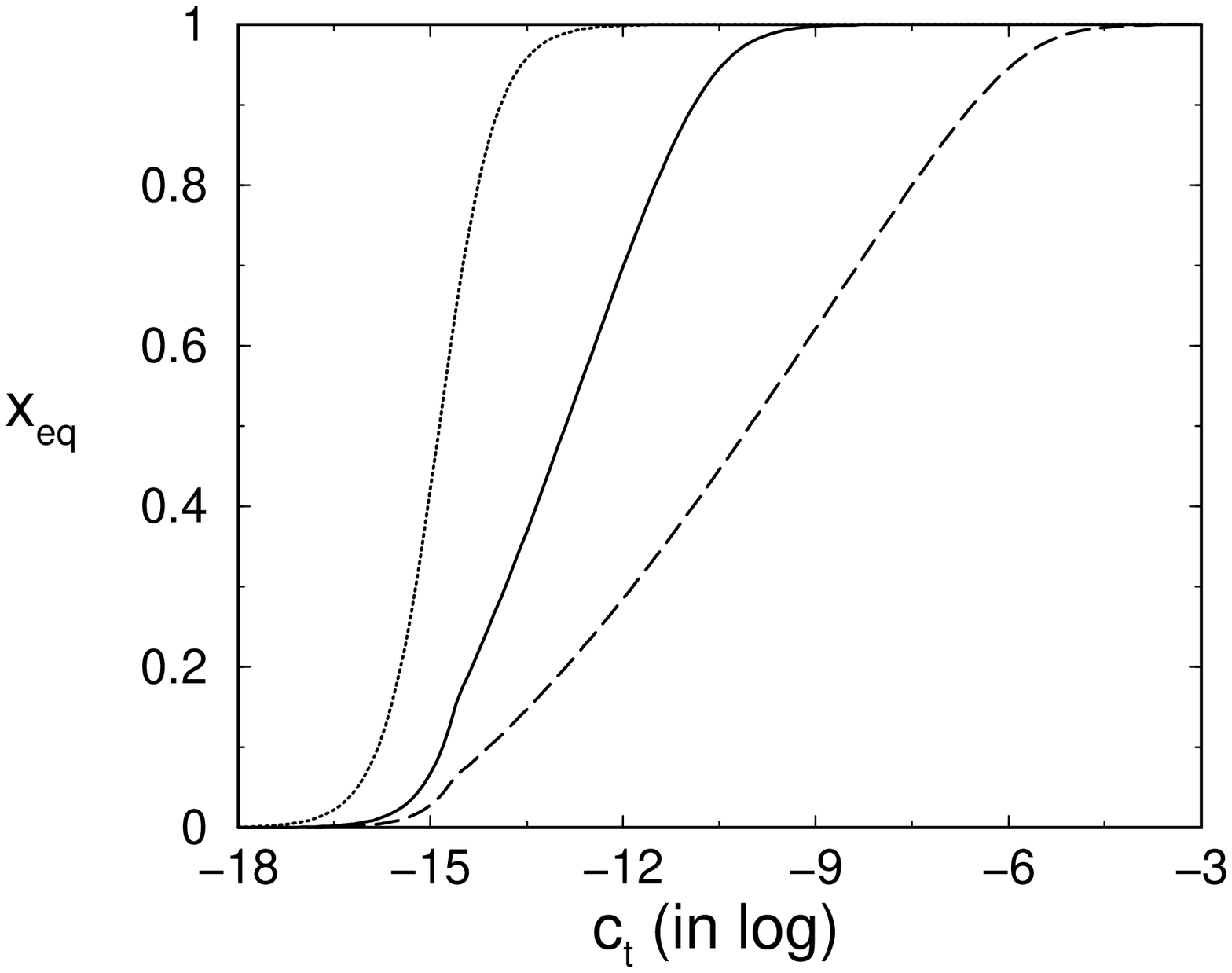}
\caption{The hybridization isotherms as calculated
using Eq.\ref{VI2b} and Eq.\ref{VI5} for the probe
target pairs utilized by Guo et al.(1994) with
$\Sigma_0 = 2500 \mathring{A}^2$ and $T = 30^{\circ}C$. 
$N = 157$ (-----), $N = 347$ (- - -) and the 
reference state case calculated from Eq.\ref{II1} 
with $N = 147$ ($\  \cdots$). The $x_{eq}$ vs $c_t$ 
curves are depicted in a) for the range $0 \leq c_t 
\leq 1 \, pM$ while $x_{eq}$ vs $\log c_t$ plots 
are depicted in b) ($-9$ corresponds to $nM$).}
\end{figure}

The complete ``long tail'' hybridization isotherm 
for the $r_F^2 < \Sigma_0 < R_F^2$ case is obtained 
from Eq.\ref{VI2b} and Eq.\ref{VI5}. In this isotherm, 
as in the interaction free Langmuir isotherm 
(Eq.\ref{II1}), $x_{eq} \rightarrow 1$ as $c_t$ 
increases. However, the two scenario differ strongly 
with respect to the range of $c_t$ involved (Fig.2).
The saturation in the long tail case occurs at a 
much higher $c_t$. When $x_{eq}$ vs $c_t$ curves
of the two scenario are compared over a limited 
$c_t$ range (Fig.2a), the long tail isotherm
is superficially similar to a Langmuir isotherm
but with apparent saturation at $x_{eq} \ll 1$. 
A plot of $x_{eq}$ vs $\log c_t$ (Fig.2b) is 
necessary in order to visualize the differences 
in the saturation behavior.

A useful measure of the sensitivity of the DNA 
chip is the $c_{50}$ corresponding to the target 
concentration, $c_{t}$, needed to obtain at
equilibrium $x_{eq} = 1/2$ (Halperin et al., 
2004a).  The $c_{50}$ also provides a rough 
estimate for the onset of saturation, as 
discussed earlier. In the $1:1$ regime, where 
the hybridization follows a Langmuir isotherm, 
$c_{50}^{1:1} = 1/K_{pt}^{1:1}$. When $R_{F}^{2} 
> \Sigma_{0} > r_{F}^{2}$, we can distinguish 
between two scenarios. So long as $x_B = 
\Sigma_{0}/R_{F}^{2} \geq 1/2$, $x_{eq} = 1/2$ 
is attained before the onset of the brush and 
$c_{50}^{1:q} = 1/K_{pt}^{1:q}$. In the opposite 
case, $x_B = \Sigma_{0}/R_{F}^{2} < 1/2$, $x_{eq} 
= 1/2$ occurs in the brush regime and $c_{50}^B 
= 1/K_{pt}^B(x_{eq} = 1/2)$. These corresponding 
experimental guidelines assume a more useful form 
when considering the logarithm of $c_{50}$. In 
particular, these relate the range of expected 
target concentrations $c_{t}$, as given by 
$c_{50}^{1:q}$ or $c_{50}^{B}$, to $\Delta 
G_{pt}^{0}$, $n$, $N$ and $\Sigma_{0}$
\begin{eqnarray}
\ln c_{50}^{1:q} & = & \frac{\Delta G_{pt}^{0}}{RT} 
+ \frac{2}{5} \ln \frac{N}{n} + N^{2/5} n 
\frac{a^{2}}{\Sigma_{0}} - n^{1/5}, \label{VI7a}\\
\ln c_{50}^{B} & = & \frac{\Delta G_{pt}^{0}}{RT} +
\frac{1}{3} \ln \frac{2 \Sigma_{0}}{n^{6/5} a^2} 
+ [ N (1-2^{2/3} x_{B}^{2/3}) + 2 n ] \left(
\frac{a^{2}}{2 \Sigma_{0}}\right)^{2/3} - n^{1/5}. 
\label{VI8}
\end{eqnarray}
$c_{50}^B$ can be significantly higher than 
$c_{50}^{1:q}$, 
\begin{equation}
\ln \frac{c_{50}^B}{c_{50}^{1:q}} = [ N (1-2^{2/3} 
x_{B}^{2/3}) + 2 n ] \left( \frac{a^{2}}{2 \Sigma_{0}}
\right)^{2/3} - N^{2/5} n \frac{a^{2}}{\Sigma_{0}} + 
\frac{1}{3} \ln \frac{2 \Sigma_{0}}{R_F^2} \gg 1,
\end{equation}
since it is dominated by the factor $\exp [ N 
(1-2^{2/3} x_B^{2/3}) (a^2/2 \Sigma_0)^{2/3}]$. 
It is helpful to compare Eq.\ref{VI7a} and Eq.\ref{VI8} 
with the Langmuir isotherm of the ``reference'' state, 
Eq.\ref{II1}, where $c_{50}^{0} = 1/K_{pt}$. The guideline 
obtained, following the same procedure, is
\begin{equation}
\ln c_{50}^{0} = \frac{\Delta G_{pt}^{0}}{RT} + 
\frac{n}{N^{4/5}} - n^{1/5}. \label{VI10}
\end{equation}
In this case $c_{50}^{0}$ is determined by $n$, $N$ and 
$\Delta G_{pt}^0/RT$. In marked contrast $c_{50}^{1:q}$ 
and $c_{50}^{B}$ depends explicitely on $\Sigma_{0}$. 
The strong $N$ dependence of $c_{50}^B$, as compared to 
$c_{50}^{1:q}$ and $c_{50}^0$, is illustrated in Fig.3. 
The increase of $c_{50}^B$ signals a corresponding loss 
of sensitivity.

\begin{figure}
\includegraphics[width=10cm]{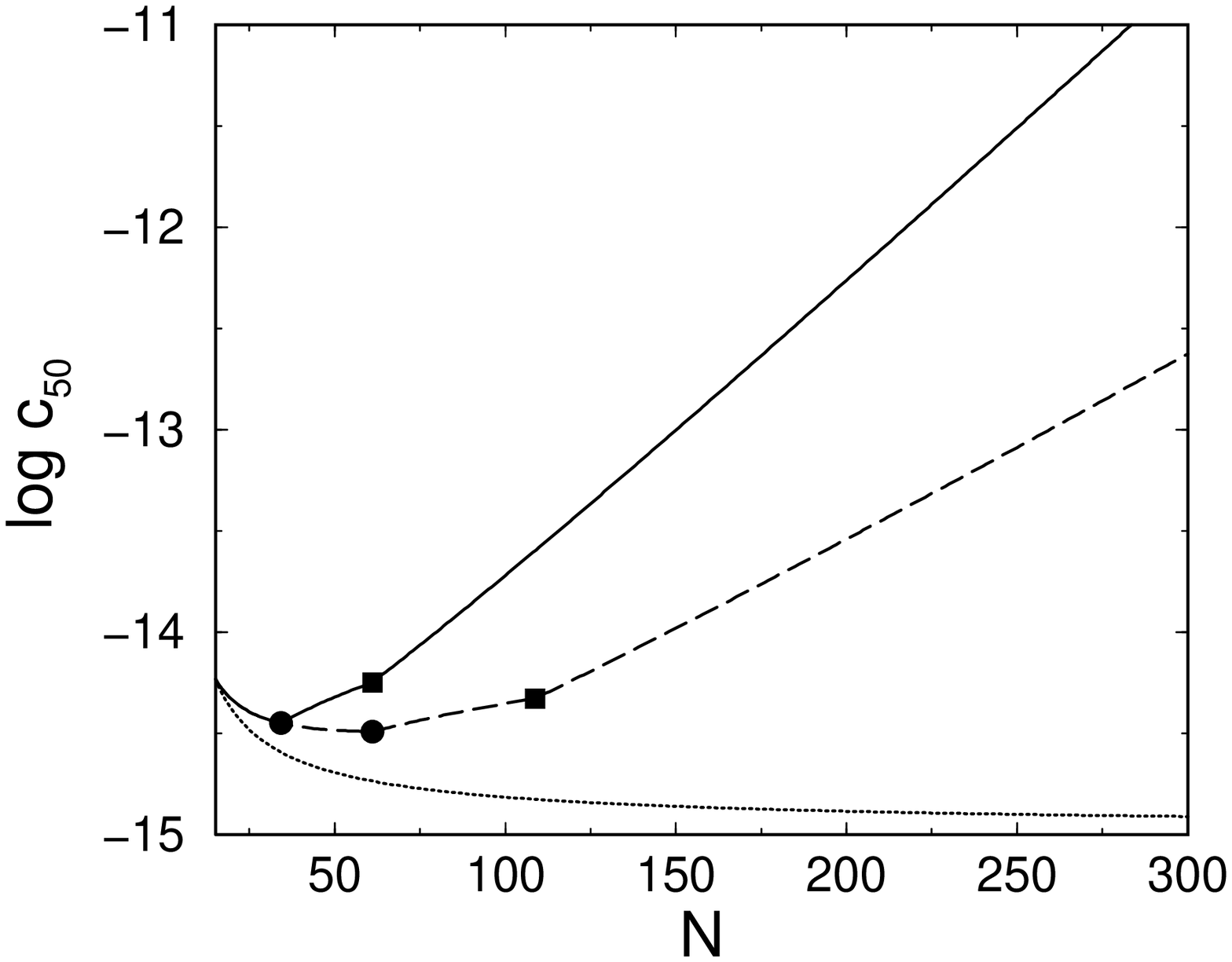}
\caption{Plots of $\log c_{50}^B$ vs $N$ for
the probes utilized by Guo et al. (1994) with
$\Sigma_0 = 2500 \mathring{A}^2$ (-----) and 
$\Sigma_0 = 5000 \mathring{A}^2$ (- - -). $T = 
30^{\circ}C$ and $n = 15$. The reference state
$\log c_{50}^0$ is plotted for comparison 
($\  \cdots$). The circles correspond to the 
crossover between $1:1$ and $1:q$ regimes whereas
squares correspond to the crossover between 
$1:q$ and $B$ regimes.}
\end{figure}

To utilize these guidelines one needs $\Delta G_{pt}^{0}$ 
as calculated using the nearest neighbor model. However,
to highlight the role of $n$ as a design parameter, it is 
helpful to use the Wetmur approximation (Wetmur, 1991) 
where average values of the nearest neighbor contributions 
are utilized. Accordingly, $\Delta G_{pt}^{0}$ of perfectly 
matched probe-target pair, when the hybridization site 
is located within the target, is approximated by
\begin{equation}
\Delta G^{0}_{pt} = (n-1) \overline{\Delta G}_{nn} +
\overline{\Delta G}_{i} + 2 \overline{\Delta G}_{e} 
\label{IV8a}
\end{equation}
where $\overline{\Delta G}_{nn}$, $\overline{\Delta G}_{i}$ 
and $\overline{\Delta G}_{e}$ are the average values 
corresponding to a nearest neighbor pair, an initiation 
step and a dangling end. Wetmur estimated the nearest 
neighbor contribution by $\overline{\Delta H}_{nn}
= -8.0 \, kcal.mol^{-1}$ and $\overline{\Delta S}_{nn} =
-21.5 \, cal.mol^{-1}.K^{-1}$, the initiation term by a
temperature independent $\overline{\Delta G}_{i} = 
2.2 \, kcal.mol^{-1}$ and the dangling end contribution 
by $\overline{\Delta H}_{e} = -8.0 \, kcal.mol^{-1}$ and
$\overline{\Delta S}_{e} = -23.5 \, cal.mol^{-1}.K^{-1}$.
Note that while useful, the Wetmur approximation 
erronously predicts identical $\Delta G_{pt}^0$ for
all $pt$ pairs with $N = n$.

\section*{BRUSH EFFECTS--KINETICS OF HYBRIDIZATION}

Having obtained the equilibrium constants $K_{pt}^{1:1}$, 
$K_{pt}^{1:q}$ and $K_{pt}^{B}(x)$ for the hybridization 
at the surface we are now in a position to consider the 
corresponding rate constants. To this end we will assume, 
and later confirm, that the rate is reaction controlled. 
Again, for simplicity, we set numerical prefactors to 
unity, $v = a^{3}$ and $l_{p} = a$. It is necessary to 
recall first the relevant features of the kinetics of 
oligonucleotide hybridization and of the desorption
of polymers out of a brush.

As discussed in Design of Oligonucleotide Microarray 
Experiments, the reference state of our analysis is 
a layer of non-interacting probes bound to a passivated 
surface by long flexible spacers. We assume that the 
molecular mechanism of hybridization in this case is 
identical to the bulk one and that the kinetics follow 
the Langmuir rate law
\begin{equation}
\frac{dx}{dt} = k_{h} c_{t} (1-x) - k_{d} x.
\label{VII1}
\end{equation}
In this regime the hybridization and denaturation 
rate constants, $k_{h}$ and $k_{d}$, are independent 
of $\Sigma_0$ or $x$ and approach their bulk values. 
At equilibrium $\frac{dx}{dt}=0$ leading to $K_{pt}= 
k_{h}/k_{d}$ as required by detailed balance. In turn, 
the hybridization mechanism of free oligonucleotides 
in solution is thought to involve the steps outlined 
below (Craig et al., 1971; P\"orschke and Eigen, 1971; 
Cantor and Schimmel, 1980; Turner, 2000). An approach
and alignment of the single stranded oligonucleotides 
is followed by the hybridization of a single base pair. 
A stable nucleus, comprising of $n_{c}+1$ base pairs, 
is formed by step-wise addition of hybridized pairs. 
Importantly, a ds sequence of $n \leq n_{c}$ is 
unstable. Once $n_{c}+1$ is attained the ds domain 
is rapidly ``zipped up''. For oligonucleotides
comprising GC base pairs $n_{c} \simeq 2-3$ and the 
hybridization rate constant exhibits the form $k_{h}= 
\tau_{h}^{-1} \exp [-\Delta G_{h}^{\#}/RT]$. Here 
$\tau_{h}$ is a molecular time scale characterizing 
the formation of the last base pair of the nucleus 
while the activation free energy $\Delta G_{h}^{\#}$ 
reflects the formation of a ds nucleus of $n_{c}$ base 
pairs plus the activation free energy for adding the next 
base pair. Importantly, the reaction is not diffusion 
controlled but involves a number of activation barriers 
associated with a corrugated free energy profile (Turner, 
2000). A rough estimate of $\Delta G_{h}^{\#}$ within the 
Wetmur approximation (Wetmur, 1991) yields $\Delta G_h^{\#} 
\simeq n_{c} \overline{\Delta G}_{nn} + \overline{\Delta 
G}_i + 2 \overline{\Delta G}_e$ indicating that $\Delta 
G_{h}^{\#}$ depends on $n_{c}$ rather than $n$. This last 
point rationalizes a phenomenological result we will 
utilize later, namely $k_h$ in high ionic strength 
solutions is 
\begin{equation}
k_{h} \simeq 10^{6} M^{-1}.s^{-1} \label{VII1b}
\end{equation}
to within one order of magnitude and with a weak $T$ 
dependence (Turner, 2000). This, together with the 
detailed balance requirement $K_{pt} = k_{h}/k_{d}$ 
yields
\begin{equation}
k_{d} \simeq 10^{6} \exp [ \Delta G_{pt}^{0}/RT ] 
\, s^{-1}. \label{VII1c}
\end{equation}
In terms of the Wetmur approximation $k_{d}$ is 
expressed as $k_{d} \simeq \tau_{h}^{-1} \exp [
(n-n_{c}) \overline{\Delta G}_{nn}/RT]$. The 
activation barrier for denaturation involves 
thus the break up of $n-n_{c}$ base pairs so as
to form an unstable ds domain. Importantly, for 
$15 \leq n \leq 25$, the denaturation life time 
at $37^{\circ}C$ is measured in years.

At this point it is of interest to comment on 
a result, obtained from computer simulations, 
concerning the kinetics of desorption out of 
a brush (Wittmer et al., 1994). It concerns a 
planar brush formed from flexible and neutral 
chains whose terminal monomer experience a short 
range attraction to the wall. The attraction was 
modeled as a well of width $a$, a monomer size, 
and depth $G_{well}$. In this system the expulsion 
rate constant is
\begin{equation}
k_{out}  = \tau^{-1}(\Sigma) \exp[
-G_{well}/RT] \label{VII5a}
\end{equation}
where $\tau(\Sigma)$ is the time required by 
the head group to diffuse across a distance 
$\Sigma^{1/2}$, corresponding to the inner 
most blob of the brush. Importantly, $k_{out}$ 
while $\Sigma$ dependent was found to be 
independent of $N$. Once the surface bond is 
broken, the expulsion of the chain out of the 
brush is driven by repulsive monomer-monomer 
interactions with neighboring chains. This last 
stage is a fast process and thus not rate 
controlling. The system studied by Wittmer et 
al. differs from ours in two respects. First, 
in this study the attractive potential is 
laterally invarient i.e., the surface is uniformly 
attractive. As a result, the reaction coordinate 
is the distance between the terminal end group and 
the surface $z$. In our case the attractive potential 
is localized at the immediate vicinity of the probe 
and the early steps of denaturation involve lateral 
separation of the two strands. Consequently the 
reaction coordinate at the vicinity of the surface 
is no longer $z$. Second, in the work of Wittmer et 
al., the barrier to adsorption is due to the brush.
There is no barrier in the mushroom regime where the 
reaction is diffusion controlled. This is also the 
case in the brush regime when the terminal group 
resides within a distance $\Sigma^{1/2}$ from the 
surface. However, as noted earlier the hybridization 
reaction in the bulk is not diffusion controlled. 
Accordingly, one should consider the possibility that 
the rate of hybridization at the surface is similarly 
not controlled by diffusion. In such a case the 
denaturation rate constant, corresponding to $k_{out}$, 
will be independent of both $N$ and $\Sigma$. 

In the following we will assume, and later 
confirm, that the rate of hybridization at
the surface is reaction controlled rather
than diffusion controlled. In quantitative 
terms, the assumption of reaction control
involves two ingredients. First, the rate 
equation may be written as
\begin{equation}
\frac{dx}{dt} = k_h c_t(z=0) (1-x) - k_d x
\end{equation}
where $c_t(z=0)$ is the local concentration
of target hybridization sites at the surface
while $k_h$ and $k_d$ stand for the rate 
constants as observed in the solution. In 
microscopic terms this implies that the 
hybridization and denaturation reactions 
at the surface are respectively monomolecular
and bimolecular and that the encounter 
probability between a probe and a target is
proportional to $c_t (z=0)$. Importantly it
also implies that the free energy surfaces of
the hybridization reaction in the bulk and at 
the surface are identical. This last point is
reasonable because this free energy surface 
reflects local reorganization of hydrogen bonds
and stacking interactions (Turner, 2000).
This assumption also implies that the lateral 
diffusion is fast enough so as to prevent 
inplane variation of $c_t (z=0)$. The second 
ingredient is the assumption that, for any $x$, 
$c_t(z=0)$ is equal to $c_t^{*} (x)$, the 
equilibrium concentration of unhybridized
terminal groups at the surface. In other words,
the diffusion of chains is sufficiently fast
in comparison to the hybridization reaction 
to ensure that a Boltzmann distribution is 
maintained. This condition is especially 
stringent in the brush regime, where inbound
diffusion must overcome a potential barrier 
due to interactions with the previously tethered
chains. The equilibrium condition requires that
$c_t^{*}/c_t = \exp ( - \Delta \mu/RT)$ where
$\Delta \mu (x)$ is the difference between the
chemical potential of a fully inserted chain
and a free one. Accordingly, for each of the 
three regimes
\begin{equation}
\frac{c_t^{*}}{c_t} = \frac{K_{pt}^{i}}{K_{pt}}
\qquad i = 1:1, 1:q, B.
\end{equation}
Note that within our treatment numerical prefactors
are omitted and there is no distinction between the
chemical potential and the free energy per chain. 
Altogether, the corresponding rate constants for
the three regimes, $i=1:1$, $1:q$ and $B$ are 
\begin{equation}
k_h^i = k_h \frac{K_{pt}^{i}}{K_{pt}}
\qquad {\rm and } \qquad k_d^i = k_d
\end{equation}
leading to
\begin{eqnarray}
k_{h}^{B} (x) & \simeq & k_{h} \left( \frac{n^{6/5} 
a^{2} x}{\Sigma_{0}} \right)^{1/3} \exp \left[ 
\frac{n}{N^{4/5}} - [ N (x^{2/3}- x_{B}^{2/3}) + 
n x^{-1/3} ] \left( \frac{a^{2}}{\Sigma_{0}}
\right)^{2/3} \right],  \label{VII2a} \\
k_{h}^{1:q} & \simeq & k_{h} \left( \frac{n}{N}
\right)^{2/5} \exp \left[ - \frac{n}{N^{4/5}} (q-1)
\right], \label{VII2b}\\
k_{h}^{1:1} & \simeq & k_{h} \left(\frac{n}{N}
\right)^{2/5}. \label{VII2c}
\end{eqnarray}
The results above where obtained assuming that the
hybridization rate is controlled by the reaction 
rather than by the diffusion towards the surface.
To check the consistency of this approach we
consider the corresponding Damk\"ohler number 
(Blanch and Clark, 1996) $Da = J_{reac}/J_{dif}$. 
Here $J_{reac}$ and $J_{dif}$ are the maximal
fluxes associated with the reaction and the inbound
diffusion, assuming reaction control. Reaction 
control implies $Da \ll 1$. $J_{reac} = k_h c_t^{*}/
\Sigma_0$ is an upper bound on the reaction flux.
The inbound flux of chain through the brush is
$J_{dif} = c_t^{*} v_{barrier}$ where $v_{barrier}$
is the diffusion velocity of a single chain at the
vicinity of the surface where the brush potential
is essentially flat. Recent experimental results
and a unified picture of theoretical models are 
presented by Titmuss et al. (2004). Altogether
\begin{equation}
Da = \frac{k_h}{\Sigma_0 v_{barrier}}
\end{equation}
where $v_{barrier} = \alpha k T/\eta N a^2$. 
Here $\eta$ is the solvent viscosity and $\alpha$
is a polymer specific numerical constant. $\alpha$
of ssDNA has not yet been determined but for 
flexible synthetic polymer $\alpha \simeq 0.1$. 
For water at $25^{\circ}C$ $\eta = 0.89 \times 
10^{-3} N.m^{-2}.s$. The Damk\"ohler number at 
$25^{\circ}C$, when both fluxes are expressed 
in units of chains$.m^{-2}.s^{-1}$ is 
\begin{equation}
Da = 0.13 \frac{N}{\Sigma_0}
\end{equation}
where we assumed $\alpha = 0.1$, $k_h = 10^6 
M^{-1}.s^{-1}$, $a = 6 \mathring{A}$ and expressed 
$\Sigma_0$ in $\mathring{A}^2$. For $100 \leq N \leq 
600$ and $T=25^{\circ}C$, the Damk\"ohler number
varies in the range $9 \times 10^{-3} \leq Da \leq
5 \times 10^{-2}$ when $\Sigma_0 = 1500 \mathring{A}^2$
and $2.6 \times 10^{-2} \leq Da \leq 0.16$ when 
$\Sigma_0 = 500 \mathring{A}^2$. The variation of 
water viscosity with temperature affects those ranges
by at most a factor $2$ for $0^{\circ} C \leq
T \leq 70^{\circ}C$. Accordingly, the assumption 
of reaction control of the hybridization rate is 
justified for typical values of $N$ and $\Sigma_0$. 
It will though fail evantualy for high $N$ values.
One should note that the issue of reaction vs 
diffusion also arise when the hybridization chamber
is agitated and we will not discuss it further.

\begin{figure}
\includegraphics[width=10cm]{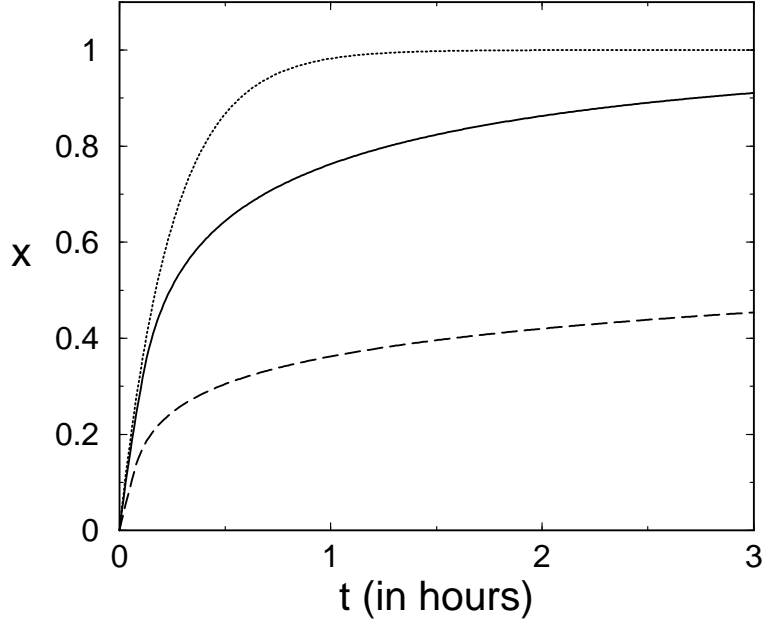}
\caption{The hybridization kinetics, as described 
by a plot of $x$ vs the time $t$ in hours, for the
probe target pairs of Guo et al. (1994): $n = 15$,
$T = 30^{\circ}C$, $c_t = 1 \, nM$ and $\Sigma_0 = 
5000 \mathring{A}^2$. The $N = 157$ (-----) and 
$N = 347$ (- - -) curves are compared to the 
reference state case with $N = 157$ ($\  \cdots$).}
\end{figure}

As required the rate constants Eq.\ref{VII2a}-\ref{VII2c} 
obey detailed balance and exhibit the proper crossover 
behavior. In particular, $k_h^i/k_d = K_{pt}^i$ as well 
as $k_h^B(x_B) = k_h^{1:q}$. The $x$ dependence of 
$k_{h}^{B}$ slows down the adsorption rate (Fig.4). 
$k_{h}^{B}(x_{co}) = k_{h}/e$ is a possible measure 
for the onset of significant slow down. In the limit 
of $N \gg 2n$ the $x^{-1/3}$ term is negligible and 
the onset occurs roughly at
\begin{equation}
x_{co} = \left[ \frac{1}{N} \left( \frac{\Sigma_{0}}{a^{2}}
\right)^{2/3} + x_{B}^{2/3} \right]^{3/2} \simeq x_{B}.
\label{VII5}
\end{equation}
It thus affects the whole brush regime. The slower kinetics 
in the brush regime can affect the attained hybridization 
even after long hybridization periods (Fig.5). This is of 
practical importance because samples of identical $c_t$ but 
different $N$ will vary in their signal intensity.

\begin{figure}
\includegraphics[width=10cm]{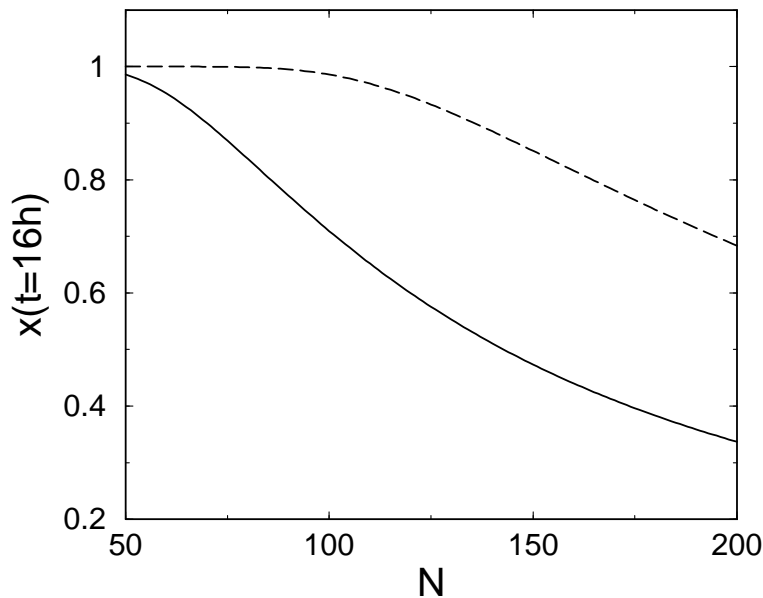}
\caption{The hybridization fraction attained after 
$t_h = 16$ hours as a function of $N$ for the Affimetrix 
probe $p2$, $n = 25$, $T = 45^{\circ}C$, $c_t = 0.1 \, 
nM$ with $\Sigma_0 = 2500 \mathring{A}^2$ (-----) and 
$\Sigma_0 = 5000 \mathring{A}^2$ (- - -).}
\end{figure}

\section*{DISCUSSION}

The relative size of the targets and probes is 
an important characteristic of oligonucleotides 
microarrays. When the two are of equal size, 
$N = n$, the onset of interaction between the 
probes is roughly set by the span of the probes 
as determined by $n$. In biology experiments 
the targets are much larger, $N \gg n$, and the 
onset of interactions is controlled by $N$. 
The progress of hybridization can give rise 
to crowding of the non-hybridized tails when
$R_{F}^{2} > \Sigma_{0}$. The polyelectrolyte 
brush thus formed affects the hybridization 
isotherm and the rate equations. In particular, 
it lowers both the hybridization rate and the 
attainable hybridization for a given concentration 
of targets. It is important to allow for this 
effect in the design of DNA microarrays, in the 
formulating of the protocols of sample preparation 
and hybridization as well as in the analysis of 
the results. With regard to design of DNA chips 
the brush effect is important in choosing the
desired density of oligonucleotide probes, or 
equivalently $\Sigma_{0}$. The brush effect will 
lower the fraction of probes that actually hybridize. 
As a result, the benefits of increasing the surface 
density of oligonucleotide probes diminish when 
the intended targets are long. When $\Sigma_{0}$ 
is set, these considerations suggest a criteria 
for tuning the length of the targets, $N$, as 
controlled by the choice of the PCR primers or 
of the fragmentation procedure. In particular, 
it is beneficial to shorten $N$ so as to avoid
crowding. When brush effects do occur the analysis 
of the results should allow for the ensuing 
deviations from the Langmuir behavior. This is 
an important point for the implementation of model 
based algorithms.

Physical chemistry type experiments, that aim 
to investigate the function of DNA microarrays, 
tend to focus on the symmetric case, of $N = n$. 
Our discussion highlights the merit of studying 
the kinetics and the equilibrium behavior in
the asymmetric case, $N \gg n$. In this case it 
is of interest to correlate the hybridization 
behavior with measurements of the brush thickness.

Our analysis focused on the case of ssDNA targets 
so as to avoid complications due to the secondary 
structure of RNA molecules. The importance of the
secondary structure of RNA targets, as used in 
gene expression experiments, is yet to be established 
because the effect of labelling by biotin is not well
understood. The effect of the fragmentation on the 
kinetics of hybridization suggests however that a 
crowding effect of some sort is indeed involved.

\section*{APPENDIX A: THE VIRIAL COEFFICIENT 
AND THE CASE OF A BRUSH OF RODS}

Consider first the second virial coefficient 
\begin{equation}
v = \frac{1}{2} \int_{0}^{\infty}\left[ 1-
\exp(-U(r)/kT) \right] 4 \pi r^{2}dr
\end{equation}
for spherical monomers of radius $a$ when their 
interactions are purely repulsive. In particular, 
the interaction potential, $U(r)$, comprises 
a hard core repulsion together with a screened 
electrostatic repulsion, that is
\begin{equation}
U = \left\{
\begin{array}{c @{ \ {\rm for} \ } c}
\infty & r < a, \\
kT {\displaystyle \frac{l_{B}}{r} \frac{\exp 
[-(r-a)/r_{D}]}{1+a/r_{D}}} &  r > a.
\end{array}
\right. \label{A1a}
\end{equation}
Here $r_{D}$ is the Debye screening length 
and $l_{B}$ is the Bjerrum length (Fowler and 
Gugenheim, 1960). The hard core contribution 
to $v$ is $2 \pi a^{3}/3$. The electrostatic 
contribution, assuming that $U/kT \ll 1$, is 
$2 \pi l_{B} r_{D}^{2}$ and altogether
\begin{equation}
v = \frac{2 \pi}{3}a^{3} + 2 \pi l_{B} r_{D}^{2}.
\label{A1b}
\end{equation}
If one supplements the electrostatic repulsion 
$U$ by a weak van der Waals attraction the first 
term assumes the form $2\pi a^{3}(1-\theta/T)/3$
where $\theta$ is the theta temperature (Rubinstein 
and Colby, 2003) thus leading to Eq.\ref{III4}. 
For $0.1 M$ of $NaCl$ salt, $r_{D} = 10 \mathring{A}$ 
and assuming $a = 6 \mathring{A}$ we find that the 
electrostatic term dominates. When the salt 
concentration is $1M$ the screening length 
diminishes to $r_{D} = 3 \mathring{A}$ and the 
two terms are comparable.  

In the case of probes and targets of equal length, 
$n=N$, the probe layer consists of a mixture of 
single stranded probes and hybridized, double
stranded ones. The associated interaction free 
energy for this case can be obtained (Halperin 
et al., 2004a) upon assuming, following Korolev 
et al. (1998), that both adopt rod-like 
configurations of equal length $L = n b$ where 
$b \simeq 3.4 \mathring{A}$ is the contribution 
of a base (base pair) to the length of the rod. 
The hybridized probes are rod-like because a 
dsDNA is rigid on the length scales of a typical 
probe ($10 < n \leq 30$). Viewing the unhybridized
probes as rigid rods is an approximation justified, 
for the short probes, by two related observations. 
One is the tendency of ssDNA to form rigid domains
of single stranded helices due to stacking 
interactions (Cantor and Schimmel, 1980; Turner, 
2000; Buhot and Halperin, 2004). The second is 
that the persistence length attributed to ssDNA 
is comparable to the 
length of the probes. It is important however to 
stress that the configurations of ssDNA are not 
yet fully characterized. As noted in Design of 
Oligonucleotide Microarray Experiments, the 
reported values of the persistence length of 
ssDNA vary over a wide range $7.5 \mathring{A} 
\leq l_{p} \leq 35 \mathring{A}$. Similarly, the
thermodynamic parameters of the stacking interactions 
are not fully established. With these reservations 
in mind, this picture provides a convenient 
approximation because it allows us to assign to 
the probe layer a unique thickness, independent 
of $x$. In particular, the thickness of the probe 
layer is comparable to $L$. The interaction free 
energy density within the probe layer is accordingly, 
$F_{int} = v c^{2}$ where $c = n(1+x)/\Sigma_0 L$ is 
the number concentration of monomers within the layer
and the interaction free energy density per unit area is
\begin{equation}
\gamma_{el} = 2 \pi l_{B} r_{D}^{2} \frac{n^{2}
(1+x)^{2}}{\Sigma_{0}^{2} L}. \label{A2}
\end{equation}
Accordingly, the overall free energy per probe site
\begin{equation}
\gamma_{site} = \gamma_{0} + x \mu_{pt}^{0} + (1-x)
\mu_{p}^{0} + \Sigma_{0} \gamma_{el}(x) + 
RT[x\ln x+(1-x)\ln(1-x)]
\label{A3}
\end{equation}
and the equilibrium condition $\mu_{t} = \mu_{pt}^{ex}
= \partial \gamma_{site}/\partial x$ yields
\begin{equation}
\frac{x_{eq}}{1-x_{eq}} = c_t K_{pt} \exp \left[  
-\Gamma(1+x_{eq}) \right]  \label{A4}
\end{equation}
with $\Gamma = 4 \pi n^{2} l_{B} r_{D}^{2}/\Sigma_{0}L$, 
as obtained earlier using the box approximation for 
the solution of the Poisson-Boltzmann equation (Halperin 
et al., 2004a) with a different prefactor. The isotherm 
obtained above differs from the ``brush isotherm'' 
because the chain elasticity does not play a role and 
the layer thickness does not exhibit an $x$ dependence.

\section*{APPENDIX B: EFFECT OF CHAIN SELF-AVOIDANCE
ON THE HYBRIDIZATION CONSTANTS}

The Flory approximation as used in the text overestimates 
both the elastic and interaction free energies. Another 
delicate point concerns the entropy of the free ends. 
At the same time, the Flory approximation is known to be
robust and its performance for the brush has been studied 
showing relatively mild deviation from the exact results 
obtained by SCF theory (Milner, 1991). With these points 
in mind it is of interest to confirm the results obtained
utilizing the Alexander-Flory approximation by a more 
rigorous approach. In the following we present exact 
results concerning $K^{1:1}$. In particular, the alternative 
derivation allows for the chain self-avoidance while ignoring 
the small correction due to interactions between the 
hybridized $ds$ domain and unhybridized tail of the target. 
To this end we utilize the partition function of a 
self-avoiding chain (Duplantier, 1989; Eisenrigler et 
al., 1982). The partition function $Z_{coil}(N)$ of a 
free self-avoiding chain of $N$ monomers is 
\begin{equation}
\label{B1}
Z_{coil}(N) = z^{N} N^{\gamma -1}
\end{equation}
where $z$ is model-dependent effective partition 
function of a monomer, and $\gamma$ is a universal 
configurational exponent. For a self-avoiding chain
with a terminal monomer anchored to an impenetrable 
planar surface, a ``mushroom'', the partition function is
\begin{equation}
\label{B2}
Z_{mush}(N) = z^{N} N^{\gamma _{1}-1} 
\end{equation}
where $\gamma _{1}$ is a different universal 
configurational exponent.

When a probe and a target hybridize, the $ds$ domain 
can be envisioned as a rigid rod with a partition 
function 
\begin{equation}
\label{B3}
Z_{rod}(n) = z_{rod}^{n} = z_{0}^{2n} \exp 
[-\Delta G_{pt}^{0}(n)/RT]
\end{equation}
Here, $z_{rod}$ is the partition function of a pair 
of hybridized monomers, $z_{0}$ is partition function 
of a single monomer in an ideal Gaussian coil, $n$ 
is number of pairs in the $ds$ domain, and $\Delta 
G_{pt}^{0}(n)$ is the free energy difference between 
the rigid $ds$ and ideal coil $ss$ domains. The free 
energy $G$ is related to partition function $Z$ by 
$G = -RT\ln (Z)$.

The hybridization constant $K_{pt}$ in a solution of 
targets and probes whose respective lengths are $N$ 
and $n \ll N$ is
\begin{equation}
\label{B4}
K_{pt} = \exp \{ - [G_{rod}(n) + G_{coil}(N-n) 
- G_{coil}(N) - G_{coil}(n)]/RT \}
\end{equation}
Using Eq.\ref{B1}, Eq.\ref{B3} and Eq.\ref{B4}, 
we obtain 
\begin{eqnarray}
\label{B5}
K_{pt} & = & \frac{Z_{rod}(n) Z_{coil}(N-n)}{Z_{coil}(n) 
Z_{coil}(N)} = \left( \frac{N-n}{N}\right)^{\gamma -1}
n^{1-\gamma} \left( \frac{z_{0}}{z}\right)^{2n} 
\exp [-\Delta G_{pt}^{0}(n)/RT] \\
\nonumber & \approx & K_{pt}^{0} n^{1-\gamma} \left( 
\frac{z_{0}}{z}\right)^{2n}
\end{eqnarray}
where $K_{pt}^{0} = \exp [-\Delta G_{pt}^{0}/RT]$ 
as introduced earlier.

For the hybridization at a surface in the $1:1$ regime 
\begin{equation}
\label{B6}
K_{pt}^{1:1} = \exp \{ - [G_{rod}(n) + G_{mush}(N-n)
-G_{coil}(N) - G_{mush}(n) -\delta S_{rod}]/RT \}
\end{equation}
Here, $\delta S_{rod}\equiv \ln (\beta )$ is the 
reduction in the rod entropy due to its attachment 
to the surface. The specific value of $\beta$ of 
the order of unity depends on the length and 
flexibility of the spacer. In the simplest case 
of a short flexible spacer, the surface eliminates 
half of space available to a free rod in the solution, 
thus yielding $\beta =1/2$. Eq.\ref{B1}, Eq.\ref{B2} 
and Eq.\ref{B6} lead to 
\begin{eqnarray}
\label{B7}
K_{pt}^{1:1} & = & \beta \frac{Z_{rod}(n) Z_{mush}
(N-n)}{Z_{mush}(n) Z_{coil}(N)} = \beta \left( 
\frac{N-n}{n}\right)^{\gamma _{1}-1} N^{1-\gamma}
\left( \frac{z_{0}}{z}\right)^{2n}
\exp [-\Delta G_{0}(n)/RT]\\
\nonumber & \approx & \beta K_{pt}^{0} N^{1-\gamma}
\left( \frac{N}{n}\right)^{\gamma_{1}-1} 
\left( \frac{z_{0}}{z}\right) ^{2n}
\end{eqnarray}
The ratio of hybridization constants at the surface 
and in solution, as determined from Eq.\ref{B5} and 
Eq.\ref{B7} is
\begin{equation}
\frac{K_{pt}^{1:1}}{K_{pt}} = \beta \left(\frac{n}{N}
\right)^{\gamma -\gamma_{1}} \qquad N \gg n
\end{equation}
The values of $\gamma \approx 1.167$ and $\gamma_{1}
\approx 0.695$ were obtained using field theoretical 
methods and numerical calculations (Duplantier, 1989; 
Eisenrigler et al., 1982). Therefore, $\gamma -
\gamma _{1} \approx 0.47$ is in close agreement with 
$K_{pt}^{1:1}/K_{pt}=(n/N)^{2/5}$, Eq.\ref{VIb}.

\begin{acknowledgments}
The authors benefitted from instructive discussions 
with E. Southern and T. Livache. EBZ was funded by 
the CEA with additional support from the Russian 
Fund for Fundamental Research (RFBR 02-0333127).
\end{acknowledgments}

\section*{REFERENCES}

Affymetrix. 2004. GeneChip Expression Analysis. 
Technical Manual. WebSite:
http://www.affymetrix.com/support/technical/manuals.affx

Alexander, S. 1977. Adsorption of chain molecules with a 
polar head - a scaling description. {\it J. Phys. (Paris)}
38:983-987.

Birshtein, T. M., Yu. V. Liatskaya, and E. B. Zhulina. 1990. 
Theory of supermolecular structure of polydisperse block 
copolymers: 1. planar layers of grafted chains. {\it Polymer}
31:2185-2196.

Birshtein, T. M., and E. B. Zhulina. 1984. Conformations
of star-branched macromolecules. {\it Polymer}, 25:1453-1461.

Bhanot, G., Y. Louzoun, J. Zhu, and C. DeLisi. 2003. The
importance of thermodynamic equilibrium for high throughput 
gene expression arrays. {\it Biophys. J.} 84:124-135.

Blanch, H. W., and D. S. Clark. 1996. Biochemical Engineering.
Marcel Dekker, New York.

Borisov, O. V., T. M. Birshtein, and E. B. Zhulina. 1991.
Collapse of grafted polyelectrolyte layer. {\it J. Phys. II.}
1:521-526.

Buhot, A., and A. Halperin. 2004. Effects of stacking on the 
configurations and elasticity of single-stranded nucleic acids.
{\it Phys. Rev. E} 70:020902(R).

Cantor, C. R., and P. R. Schimmel. 1980. Biophysical Chemistry. 
WH Freeman, New York.

Chan, V., D. J. Graves, and S. McKenzie. 1995. The biophysics 
of DNA hybridization with immobilized oligonucleotide probes.
{\it Biophys. J.} 69:2243-2255.

Craig, M. E., D. M. Crothers, and P. Doty. 1971. Relaxation 
kinetics of dimer formation by self complementary 
oligonucleotides. {\it J. Mol. Biol.} 62:383-392.

Dai, H., M. Meyer, S. Stepaniants, M. Ziman, and R. Soughton.
2002. Use of hybridization kinetics for differentiating 
specific from non-specific binding to oligonucleotide 
microarrays. {\it Nucleic Acids Res.} 30:e86.

De Gennes, P. G. 1979. Scaling Concepts in Polymer
Physics. Cornell University Press, Ithaca, New York, USA.

DiMarzio, E. A. 1965. Proper Accounting of Conformations
of a Polymer Near a Surface. {\it J. Chem. Phys.} 42:2101-2106.

Duplantier, B. 1989. Statistical Mechanics of Polymer 
Networks of Any Topology. {\it J. Stat. Phys.} 54:581-680.

Eisenriegler, E., K. Kremer, and K. Binder 1982. 
Adsorption of Polymer Chains at Surfaces: Scaling 
and Monte Carlo Analyses. {\it J. Chem. Phys.} 
77:6296-6320.

Evans, D. F., and H. Wennerstr\"om. 1994. The Colloid 
Domain. VHC, New York.

Forman, J. E., I. D. Walton, D. Stern, R. P. Rava, and 
M. O. Trulson. 1998. Thermodynamics of duplex formation 
and mismatch discriminiation on photolithographically 
synthesized oligonucleotide arrays. {\it ACS Symp. 
Ser.} 682:206-228.

Fowler, R. and E. A. Gugenheim. 1960. Statistical 
Thermodynamics. Cambridge, UK.

Georgiadis, R., K. P. Peterlinz, and A. W. Peterson. 
2000. Quantitative Measurements and Modeling of Kinetics 
in Nucleic Acid Monolayer Films Using SPR Spectroscopy. 
{\it J. Am. Chem. Soc.} 122:3166-3173.

Graves, D. J. 1999. Powerful tools for genetic analysis 
come of age. {\it Trends Biotechnol.} 17:127-134.

Guo, Z., R. A. Guilfoyle, A. J. Thiel, R. Wang, and 
L. M. Smith. 1994. Direct Fluorescence Analysis of 
Genetic Polymorphism by Hybridization with Oligonucleotide 
Arrays on Glass Supports. {\it Nucleic Acids Res.} 
22:5456-5465.

Halperin, A., M. Tirrell, T. P. Lodge. 1992. Tethered Chains 
in Polymer Microstructures. {\it Adv. Polym. Sci.} 100:31-71.

Halperin, A., A. Buhot, and E. B. Zhulina. 2004a. Sensitivity, 
specificity and the hybridization isotherms of DNA chips. 
{\it Biophys J.} 86:718-730.

Halperin A., A. Buhot, and E. B. Zhulina. 2004b. Hybridization 
Isotherms of DNA Chips and the Quantification of Mutation Studies. 
{\it Clin. Chem.} 50:2254-2262.

Hekstra, D., A. R. Taussig, M. Magnasco, and F. Naef. 2003.
Absolute mRNA concentrations from sequence-specific 
calibration of oligonucleotide arrays. {\it Nucleic Acids 
Res.} 31:1962-1968.

Held, G. A., G. Grinstein, and Y. Tu. 2003. Modeling of DNA 
microarray data by using physical properties of hybridization. 
{\it Proc. Natl. Acad. Sci. USA} 100:7575-7580.

Heller, M. J. 2002. DNA microaaray technology: devices, systems
and applications. {\it Annu. Rev. Biomed. End} 4:129-153.

HyTer$^{TM}$ version 1.0, Nicolas Peyret and John SantaLucia Jr. 
Wayne State University. 
http://ozone2.chem.wayne.edu/Hyther/hytherm1main.html

Kepler, T. B., L. Crosby, and K. T. Morgan. 2002. Normalization 
and analysis of DNA microarray data by self consistency and local 
regression. {\it Genome Biol.} 3:0037.1-0037.12.

Korolev, N., A. P. Lyubartsev, and L. Nordenski\"old. 1998.
Application of Polyelectrolyte Theories for Analysis of
DNA Melting in the Presence of $Na^{+}$ and $Mg^{2+}$ Ions.
{\it Biophys. J.} 75:3041-3056.

Livshits, M. A., and A. D. Mirzabekov. 1996. Theoretical 
analysis of the kinetics of DNA hybridization with
gel-immobilized oligonucleotides. {\it Biophys. J.} 71:2795-2801.

Lockhart, D. J., and E. A. Winzeler. 2000. Genomics, gene 
expression and DNA arrays. {\it Nature} 405:827-836.

Mills, J. B., E. Vacano, and P. J. Hagerman. 1999. 
Flexibility of Single-stranded DNA: Use of Gapped Duplex
Helices to Determine the Persistence Lengths of Poly(dT) 
and Poly(dA). {\it J. Mol. Biol.} 285:245-257.

Milner, S. T. 1991. Polymer brushes. {\it Science} 251:905-914.

Moore, W. J. 1972. Physical Chemistry. Longman, London, UK.

Naef, F., M. Magnasco. 2003. Solving the riddle of the bright 
mismatches: Labeling and effective binding in oligonucleotide 
arrays. {\it Phys. Rev. E} 68:011906.

Nelson, B. P., T. E. Grimsrud, M. R.  Liles, R. M. Goodman, and 
R. M. Corn. 2001. Surface Plasmon Resonance Imaging Measurements 
of DNA and RNA Hybridization Adsorption onto DNA Microarrays. 
{\it Anal. Chem.} 73:1-7.

Okahata, Y., M. Kawase, K. Niikura, F. Ohtake, H. Furusawa, and 
Y. Ebara. 1998. Kinetic Measurements of DNA Hybridization on an 
Oligonucleotide-Immobilized 27-MHz Quartz Crystal Microbalance.
{\it Anal. Chem.} 70:1288-1296.

Peterson, A. W., L. K. Wolf, and R. M. Georgiadis. 2002. 
Hybridization of Mismatched or Partially Matched DNA at Surfaces. 
{\it J. Am. Chem. Soc.} 124:14601-14607.

Peyret, N., P. A. Seneviratne, H. T. Allawi, and J. SantaLucia.
1999. Nearest-neighbor thermodynamics and NMR of DNA sequences
with internal A.A, C.C, G.G, and T.T mismatches. 
{\it Biochemistry} 38:3468-3477.

Pincus, P. 1991. Colloid stabilization with grafted 
polyelectrolytes. {\it Macromolecules} 24:2912-2919.

Pirrung, M. C. 2002. How to make a DNA chip? {\it Angew. 
Chem. Int. Ed.} 41:1277-1289.

P\"{o}rschke, D., and M. Eigen. 1971. Co-operative non-enzymatic 
base recognition III. Kinetics of the helix--coil transition of
the oligoribouridylic$\cdot$oligoriboadenylic acid system and of
oligoriboadenylic acid alone at acidic pH. {\it J. Mol. Biol.} 
62:361-364.

Prix, L., P. Uciechowski, B. B\"{o}ckmann, M. Giesing, and A. J. 
Schuetz. 2002. Diagnostic Biochip Array for Fast and Sensitive 
Detection of K-{\it ras} Mutations in Stool. {\it Clin. Chem.} 
48:428-435.

Rosenow, C., R. M. Saxena, M. Durst, and T. R. Gingeras. 2001.
Prokaryotic RNA preparation methods useful for high density 
array analysis: comparison of two approaches. {\it Nucleic Acids 
Res.} 29:e112.

Rubinstein, M., and R. H. Colby. 2003. Polymer Physics, Oxford 
University Press, Oxford.

R\"{u}he, J., M. Ballauff, M. Biesalski, P. Dziezok, F. Grohn, 
D. Johannsmann, N. Houbenov, N. Hugenberg, R. Konradi, S. Minko,
M. Motornov, R. R. Netz, M. Schmidt, C. Siedel, M. Stamm, 
T. Stephan, D. Usov, and H. N. Zhang. 2004. Polyelectrolyte Brushes. 
{\it Adv. Polym. Sci.} 165:77-150.

SantaLucia, J. 1998. A unified view of polymer, dumbbell, and 
oligonucleotide DNA nearest-neighbor thermodynamics. {\it Proc.
Natl. Acad. Sci. USA} 95:1460-1465.

SantaLucia, J., and D. Hicks. 2004. The Thermodynamics of DNA 
Structural Motifs. {\it Annu. Rev. Biophys. Biomol. Struct.} 
33:415-450.

Smith, S. B., Y. J. Cui, and C. Bustamante. 1996. Overstretching
B-DNA: The elastic response of individual double-stranded and 
single-stranded DNA molecules. {\it Science} 271:795-799.

Southern, E., K. Mir, and M. Shchepinov. 1999. Molecular interactions 
on microarrays. {\it Nat. Genet.} 21:5-9.

Steel, A. B., T. M. Herne, and M. J. Tarlov. 1998. Electrochemical
Quantitation of DNA Immobilized on Gold. {\it Anal. Chem.} 
70:4670-4677.

Strick, T. R., M.-N. Dessinges, G. Charvin, N. H. Dekker, J.-F. 
Allemand, D. Bensimon, and V. Croquette. 2003. Stretching of
macromolecules and proteins. {\it Rep. Prog. Phys.} 66:1-45.

Su, H.-J., S. Surrey, S. E. McKenzie, P. Fortina, and D. J. 
Graves. 2002. {\it Electrophoresis} 23:1551-1557.

Titmuss, S., W. H. Briscoe, I. E. Dunlop, G. Sakellariou, N.
Hadjichristidis, and J. Klein. 2004. Effect of end-group 
sticking energy on the properties of polymer brushes: 
Comparing experiment and theory. {\it J. Chem. Phys.} 
121:11408-19.

Turner, D. H. 2000. Chap.8 Conformational Changes. in Bloomfield, 
V. A., D. Crothers, and I. Tinoco Jr. Nucleic Acids: structures, 
properties and functions. University Science Books, Sausalito, USA. 

Vainrub, A., and M. B. Pettitt. 2002. Coulomb blockage of 
hybridization in two-dimensional DNA arrays. {\it Phys. Rev. E}
66:041905.

Wang, J. 2000. From DNA biosensors to gene chips. {\it Nucleic 
Acids Res.} 28:3011-3016.

Wetmur, J. G. 1991. DNA Probes: Applications of the Principles 
of Nucleic Acid Hybridization. {\it Crit. Rev. Biochem. Mol. 
Bio.} 26:227-259. 

Wittmer, J., A. Johner, J. F. Joanny, and K. Binder. 1994.
Chain desorption from semidilute polymer brush: A Monte Carlo
simulation. {\it J. Chem. Phys.} 101:4379-4390.

Zhang, L., M. F. Miles, K. D. Aldape. 2003. A model of molecular
interactions on short oligonucleotide microarrays. {\it Nature 
Biotechnol.} 21:818-821.

\end{document}